\documentclass[preprint,12pt]{elsarticle}
\usepackage{amsmath}
\usepackage{amsfonts}
\usepackage{bm}
\usepackage{lipsum}
\usepackage{enumerate}
\usepackage{graphicx}
\usepackage{amsfonts}

\usepackage{mathrsfs}
\usepackage{subfigure}
\usepackage{epstopdf}
\usepackage{url}
\usepackage{verbatim}
\usepackage{amssymb}
\usepackage{hyperref}
\usepackage{color}
\usepackage{listings}

\def\@begintheorem#1#2{\par\bgroup{\scshape #1\ #2. }\it\ignorespaces}
\def\@opargbegintheorem#1#2#3{\par\bgroup%
   {\scshape #1\ #2\ ({\upshape #3}). }\it\ignorespaces}
\def\@endtheorem{\egroup}

\if@onethmnum
  \newtheorem{theorem}{Theorem}
  \newtheorem{lemma}[theorem]{Lemma}
  \newtheorem{corollary}[theorem]{Corollary}
  \newtheorem{proposition}[theorem]{Proposition}
  \newtheorem{definition}[theorem]{Definition}
\newtheorem{example}[theorem]{Example}
\newtheorem{remark}[theorem]{Remark}
\newtheorem{homework}[theorem]{Homework}
\newtheorem{case}[theorem]{}

\else

\fi

\usepackage[T1]{fontenc}

\journal{Physical Review Fluids }

\begin{document}
%\maketitle
%\tableofcontents

\begin{frontmatter}

\title{Dispersion induced by unsteady diffusion-driven flow in parallel-plate channel}

\author[1,2]{Lingyun Ding}
\ead{ dingly@g.ucla.edu}
  \address[1]{Department of Mathematics, University of California Los Angeles, CA, 90095, United States}
  
\author[2]{Richard M. McLaughlin  \corref{mycorrespondingauthor}}
\ead{rmm@email.unc.edu}
\cortext[mycorrespondingauthor]{Corresponding author}
\address[2]{Department of Mathematics, University of North Carolina, Chapel Hill, NC, 27599, United States}
%\date{August 2022}

\begin{abstract}
We investigate diffusion-driven flows in a parallel-plate channel domain with linear density stratification, which arise from the combined influence of gravity and diffusion in density-stratified fluids. We compute the time-dependent diffusion-driven flows and perturbed density field using eigenfunction expansions under the Boussinesq approximation. In channel domain, the unsteady flow converges to a steady-state solution either monotonically or non-monotonically  (highly oscillatory), depending on the relation between the Schmidt number and the non-dimensionalized stratified scalar diffusivity, while the flow in the half-space inclined plane problem exhibits oscillatory convergence for all parameters. To validate the Boussinesq approximation, we propose the quasi-Boussinesq approximation, which includes transverse density variation in the inertial term. Numerical solutions show that the relative difference between the Boussinesq and quasi-Boussinesq approximations is uniformly small. We also study the mixing of a passive tracer induced by the advection of the unsteady diffusion-driven flow and present the series representation of the time-dependent effective diffusion coefficient. For small Schmidt numbers, the effective diffusion coefficient induced by the unsteady flow solution can oscillate with an amplitude larger than the effective diffusion coefficient induced by the long-time-limiting steady-state flow.  Interestingly, the unsteady flow solution can reduce the time-dependent effective diffusion coefficient temporally in some parameter regimes, below even that produced by pure molecular diffusion in the absence of a flow. However, at long times, the effective diffusion is significantly enhanced for large P\'eclet numbers.
\end{abstract}

%35Q30  	Navier-Stokes equations
% 82C70Transport processes
% 82C80 Numerical methods (Monte Carlo, series resummation, etc.)
%34E13 Multiple scale methods
%76R50 diffusion
%37A25: Ergodicity, mixing, rates of mixing
%37H10: Generation, random and stochastic difference and differential equations
%37N10: Dynamical systems in fluid mechanics, oceanography and meteorology
  
\end{frontmatter}

\section{Introduction}

Diffusion-driven flow is a boundary layer flow that results from the combined influence of gravity and diffusion, which exists in the density-stratified fluids whenever the gravity field is not parallel to the solid boundary. The hydrostatic equilibrium in the density-stratified fluid with diffusive solute requires two conditions. First, isopycnals should be perpendicular to the direction of gravity. Second, the impermeable (i.e. no-flux) boundary condition requires that the isopycnals must always be perpendicular to an impermeable boundary to ensure that there is no diffusive flux normal to the boundary. Therefore, when the impermeable boundary is not parallel to the direction of gravity, isopycnals can not be perpendicular to both of them at the same time. The breaking of the hydrostatic equilibrium yields the diffusion-driven flow. 

The diffusion-driven flow is at the same scale as molecular diffusion due to the formation mechanism, and as such could lead to interesting dynamics on long time scales or on small length scales. Therefore, the study of diffusion-driven flow historically was motivated by understanding the transport and mixing over geological time scales such as the ocean boundary mixing \cite{phillips1970flows,wunsch1970oceanic} and salt transport in rock fissures \cite{woods1992natural,heitz2005optimizing,zagumennyi2016diffusion,shaughnessy1995low}. The recent applications of diffusion-driven flow have been expanded in many areas. The potential high-impact studies include the self-propulsion of immersed objects \cite{mercier2014self,allshouse2010propulsion}, the molecular diffusivity measurement \cite{allshouse2010novel}, the self-assembly of particles in a stratified fluid \cite{camassa2019first} and airflows created by mushrooms for dispersing their spores \cite{dressaire2016mushrooms}.

We find two points that have not been addressed well in the literature. First, the studies mentioned above mainly concern the long-time stationary configuration of the diffusion-driven flow, but little is known about the transient dynamics at the earlier stage of the diffusion-driven flow formation, which can  play an important role in some parameter regimes.  Kistovich et al. \cite{kistovich1993structure} studied the transient diffusion-driven flow induced by the inclined plane using Fourier series expansion. The series representation of the flow converges rapidly at fixed time, but suffers from non-uniformity in time as the truncations are all unbounded as time grows.  Harabin \cite{harabin2016diffusively} presented a different perspective of the same problem. He derived the flow solution valid for all time scales using the Laplace transform and demonstrate that the flow exhibits oscillatory behavior in its evolution for small Schmidt (Prandtl) numbers.

Hence, the first goal of this study is to generalize those results to tilted parallel-plate channel domain and to show how the flow properties change due to the boundary geometries.  We explicitly calculate the time-dependent flow solution and the perturbed density field starting from a uniform linear density stratification using the eigenfunction expansion.  Interestingly, for the channel case, the time-dependent diffusion-driven flow exhibits oscillations for some parameters and decays monotonically for other parameter combinations. This is different from the flows in the inclined plane problem, which always includes oscillating terms.

Second, the evolution of a passive scalar is crucial in numerous fields including microfluidics \cite{aminian2016boundaries,stroock2002chaotic}, biology \cite{lin2011stirring,lin2022stirring}, and oceanography\cite{thomas2022wave}. 
Using the steady diffusion-driven flow solution, \cite{woods1992natural,heitz2005optimizing} studied the optimal gap thickness and angle to maximize long time mixing of a passive scalar advected by a steady flow arising in the tilted parallel-plate channel domain. Intuitively, unsteady diffusion-driven flows generate different properties than their steady counterparts, while they are less studied in the literature, and investigating them is the second goal of this work.  Using the time-dependent flow formula we derived, we calculate the effective diffusion coefficient of the passive scalar, which is a fundamental quantity to characterize the passive scalar distribution.  Similar as in the steady case, the unsteady diffusion-driven flow solution could significantly enhance the tracer dispersion. However, in some parameter regimes, the unsteady flow solution introduces considerably large oscillations in the effective diffusion coefficient and can even decrease the mixing coefficient temporally.

This paper is organized as follows. In section \ref{sec:diffusion induce flow Governing equation and non dimensionalization}, we formulate the governing equation for the diffusion-driven flow and document the non-dimensionalization  procedure. In section \ref{sec:Flow equation}, we derive the expression of the diffusion-driven flow and the coupled density perturbation. In section \ref{sec:Effective diffusivity}, we study the effective diffusion coefficient of the passive scalar induced by the diffusion-driven flow and explore the optimal parameters for enhancing or reducing the effective diffusion coefficient.

\section{Governing equation and nondimensionalization }
\label{sec:diffusion induce flow Governing equation and non dimensionalization}

\subsection{Governing equation}
\begin{figure}
  \centering
    \includegraphics[width=0.46\linewidth]{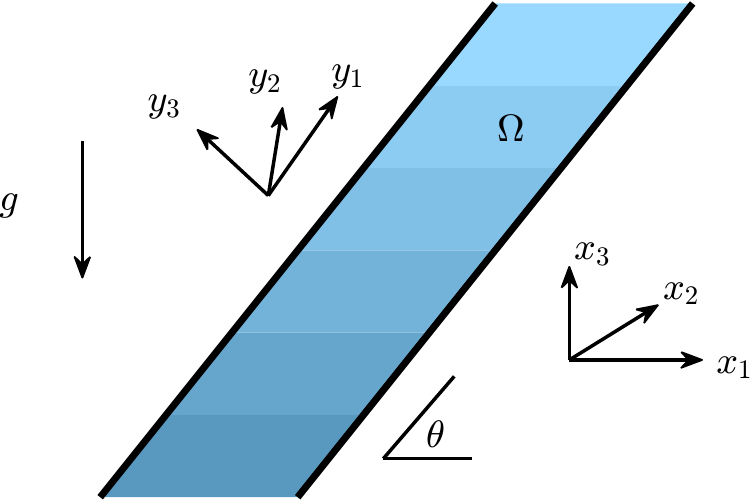}
  \hfill
  \caption[Schematic showing the setup for the diffusion-driven flow problem]
  { Schematic showing the setup for the diffusion-driven flow problem. }
  \label{fig:DiffusionDrivenFlowSchematic}
\end{figure}

There could be two different types of scalars in the system we consider: the stratifying scalar, $C$ and a passive scalar $T$. The stratified scalar contributes to the density stratification, which creates diffusion-driven flows. The system could also include passive scalars, such as a fluorescent dye. The passive scalar will be passively advected by the fluid flow without changing the velocity field.  Both scalars satisfy the advection-diffusion equation with no-flux boundary conditions, and the equation for the passive scalar takes the form
\begin{equation}\label{eq:Advection Diffusion Equation}
\partial_{t}T+ \mathbf{u}(\mathbf{x},t)\cdot \nabla T=  \kappa_{p} \Delta T, \quad T(x,\mathbf{y},0)= T_{I}(x,\mathbf{y}),\quad  \left. \partial_{\mathbf{n}} T\right|_{boundary}=  0,
\end{equation}
where $\kappa_{p}$ is the passive scalar diffusivity, $T_{I}(x,\mathbf{y})$ is the initial data, $\mathbf{n}$ is the outward normal vector of the boundary.
Figure \ref{fig:DiffusionDrivenFlowSchematic} sketches two coordinate systems for a tilted parallel-plate channel domain with a inclination angle $\theta$ which  satisfies $0\leq \theta \leq \frac{\pi}{2}$. In this setup, $x_3$-direction is parallel to the direction of gravity, $y_1$-direction is the longitudinal direction of the channel. $\Omega=\left\{ y_{3}| y_3 \in [0,L] \right\}$ is the cross-section of the channel.  The relation between  the lab frame coordinates $(x_{1},x_{2},x_{3})$ and the coordinates $(y_{1},y_{2},y_{3})$ is 
\begin{equation}
\begin{aligned}
  \begin{bmatrix}
    y_{1}\\
    y_{3}\\
  \end{bmatrix}
  &=
  \begin{bmatrix}
    \cos \theta &\sin \theta \\
    -\sin \theta &\cos \theta\\
  \end{bmatrix}
  \begin{bmatrix}
    x_{1}\\
    x_{3}\\
  \end{bmatrix}, \quad y_{2}=x_{2}.
\end{aligned}
\end{equation}
In $(y_{1},y_{2},y_{3})$ coordinates system, the direction of gravity is  $(-\sin \theta, 0, -\cos \theta)$.

We assume the fluid density linearly depends on the stratified scalar. For example, the density of sodium chloride solution increases linearly as the concentration increases \cite{hall1924densities}.  Therefore, the density field $\rho$ and the fluid flow $u_{i}$ satisfies the incompressible Navier-Stokes equation,
\begin{equation}
\begin{aligned}
&\rho \left(  \partial_{t}  u_{i}+  \mathbf{u}\cdot \nabla  u_{i} \right) = \mu \Delta u_{i} - \partial_{x_{i}} p- \rho g \delta_{i3},\quad  \left. u_{i} \right|_{ \partial \Omega }=0,\quad i=1,2,3, \quad\nabla \cdot \mathbf{u}=0,\\
& \partial_{t}\rho+ \mathbf{u}\cdot \nabla \rho= \kappa_{s} \Delta\rho,\quad \left. \partial_{\mathbf{n}}\rho \right|_{\partial \Omega }=0, \quad \left. \rho \right|_{|x_{3}| \rightarrow \infty }=\rho_{0}-\Gamma x_{3},\\
\end{aligned}
\end{equation}
where $\delta_{ij}$ is the Kronecker delta, $g (cm/s^{2})$  is the acceleration of gravity, $\Gamma$ $(gram \cdot cm^{-4})$ is the density gradient, $\mu$, $ (gram \cdot cm^{-1} \cdot s^{-1})$ is the dynamic viscosity,  $p$ $(gram\cdot cm^{-1}\cdot s^{-2})$ is the pressure and $\kappa_s$ $(cm^{2}/s)$ is the molecular diffusivity of the stratified scalar. In this study, we make the assumption that the background density function varies linearly with height. This assumption is a local approximation to the scenario where the density function changes slowly with respect to height. By assuming this linearity, we aim to simplify the analysis while still capturing the essential behavior of the system.

\subsection{Nondimensionalization }
Since we are interested in the dispersion of the passive scalar, we use the diffusion time scale of the passive scalar as the characteristic time scale of the whole system. With the change of variables
\begin{equation}
\begin{aligned}
  &\rho_{0}\rho'=\rho , \quad\frac{L^2}{\kappa_{p}} t' =t, \quad Lx'=x, \quad Uu'=u, \quad \frac{\mu U}{L} p' =p, \quad  \kappa_{p}\kappa'=\kappa,\quad \frac{\rho_0}{L}\Gamma_{0}=\Gamma,\\
 &T' (\mathbf{x}',t') L^{-3}\int\limits_{\mathbb{R}\times \Omega}^{}T_{I} (\mathbf{x})\mathrm{d}\mathbf{x} =T (\mathbf{x},t), 
\end{aligned}
\end{equation}
we have
\begin{equation}
  \begin{aligned}
&\frac{\rho_0U \kappa_{p}}{L^2}  \rho' \partial_{t'}  u'_{i}+ \frac{\rho_0U^{2}}{L} \rho' \mathbf{u}' \cdot \nabla_{\mathbf{x}'}u'_{i}  =\frac{ \mu U}{L^2} \Delta_{\mathbf{x}'}u'_{i} - \frac{\mu U}{L^2}\partial_{x'_{i}} p'- \rho_0g \rho'  \delta_{i3},\; i=1,2,3,  \\
& \frac{L^2}{\kappa_{p}}  \partial_{t'}T'+ \frac{U}{L}\mathbf{u} \nabla_{\mathbf{x}'}T'= \frac{ \kappa_{p}}{L^{2}} \Delta_{\mathbf{x}'} T',\\
&\frac{\rho_0\kappa_{p}}{L^2} \partial_{t'}\rho'+ \frac{U \tilde{\rho}}{L} \mathbf{u}' \cdot \nabla _{\mathbf{x}'}\rho'= \frac{\kappa_{s} \tilde{\rho}}{L^2} \Delta_{\mathbf{x}'}\rho'.\\
\end{aligned}
\end{equation}
We can drop the primes without confusion and obtain the nondimensionalized version
\begin{equation}\label{eq:NS nondimensional 3D diffusion driven flow}
\begin{aligned} 
&\frac{\mathrm{Re}}{ \mathrm{Pe_{p}} }  \rho \partial_{t}  u_{i}+ \mathrm{Re} \rho \mathbf{u} \cdot \nabla  u_{i}  =\Delta u_{i} - \partial_{x_{i}} p- \frac{\mathrm{Re}}{\mathrm{Fr^{2}}}\rho  \delta_{i3}, \quad i=1,2,3,\\
& \partial_{t}T+ \mathrm{Pe}_{p}  \mathbf{u} \cdot \nabla T= \Delta T,\\
&\frac{1}{ \kappa_{2}}\partial_{t}\rho+ \mathrm{Pe}_{s}  \mathbf{u} \cdot \nabla \rho= \Delta\rho,\\
\end{aligned}
\end{equation}
where the non-dimensional parameters are the non-dimensionalized stratified scalar diffusivity $\kappa_{2}=\frac{\kappa_{s}}{\kappa_{p}}$, P\'eclet number $\mathrm{Pe}_{s}= \frac{UL}{\kappa_{s}}$ and $\mathrm{Pe}_{p}= \frac{UL}{\kappa_{p}}$, Reynolds number $\mathrm{Re}= \frac{\rho_{0} L U}{\mu}$, Froude number $\mathrm{Fr}= \frac{U}{\sqrt{gL}}$, and Schmidt number $\mathrm{Sc}= \frac{\mu}{\rho_{0} \kappa_{p}}=\frac{\mathrm{Pe}_{p}}{\mathrm{Re}}$. If the scalar field is the temperature field, then $\kappa_{p}$ is the thermal diffusivity and $\frac{\mu}{\rho_{0} \kappa_{p}}=\frac{\mathrm{Pe}_{p}}{\mathrm{Re}}$ is the Prandtl number.

It is convenient to consider the problem in $(y_{1},y_{2},y_{3})$ coordinate system. We denote $v_i$ as the velocity component along the $y_{i}$-direction. Since the initial condition and the boundary condition are independent of $y_2$, equation \eqref{eq:Advection Diffusion Equation} and \eqref{eq:NS nondimensional 3D diffusion driven flow}  becomes
\begin{equation}\label{eq:NS nondimensional 3D diffusion driven flow rotated coordinate}
  \begin{aligned}
&\rho \left( \frac{1}{ \mathrm{Sc}}   \partial_{t}  v_{1}+ \mathrm{Re} v_{1}\partial_{y_{1}} v_{1} + \mathrm{Re} v_{3}\partial_{y_{3}} v_{1}\right) =\Delta v_{1} - \partial_{y_{1}} p- \frac{\mathrm{Re}}{\mathrm{Fr^{2}}}\rho  \sin \theta,  \\
&\rho \left( \frac{1}{ \mathrm{Sc}}  \partial_{t}  v_{3}+ \mathrm{Re} v_{1}\partial_{y_{1}} v_{3} +\mathrm{Re}  v_{3}\partial_{y_{3}} v_{3}\right) = \Delta v_{3} - \partial_{y_{3}} p- \frac{\mathrm{Re}}{\mathrm{Fr^{2}}}\rho  \cos \theta,  \\
& \partial_{t}T+ \mathrm{Pe}_{p}  \mathbf{v} \cdot \nabla T= \Delta T,\\
&\frac{1}{ \kappa_{2}}\partial_{t}\rho+ \mathrm{Pe}_{s}  \mathbf{v} \cdot \nabla \rho= \Delta\rho,\quad \left. \rho \right|_{|\mathbf{y}|\rightarrow \infty }=\rho_0- \Gamma_0 (y_{1}\sin \theta + y_{3} \cos \theta).\\
\end{aligned}
\end{equation}

We next consider some combination of experimental physical parameters, which could give us the order of magnitude of the non-dimensional parameters and help with the perturbation analysis. The scaling relation for the characteristic velocity and the physical parameter varies for different boundary geometries. According to the formula in \cite{phillips1970flows,heitz2005optimizing},  the characteristic velocity of steady diffusion-driven flow in the parallel-plate channel is $U =\kappa \left( \frac{ g  \Gamma }{\mu \kappa} \right)^{\frac{1}{4}}$ and the characteristic boundary layer thickness is $L_{b}=\left( \frac{ g  \Gamma }{\mu \kappa} \right)^{-\frac{1}{4}}$. In an experiment with sodium chloride solution, the parameters could be  $g=980$  cm/s$^{2}$, $\mu=0.01$ gram/(cm.s), $\kappa_{s}= 1.5\times10^{-5}$ cm$^{2}$/s, $\Gamma=0.007$ gram/$cm^{4}$, $\rho=1$ gram/cm$^{3}$, we have $U =0.00123353$ cm/s, $L_{b}= 0.0121602$ cm. If $L=0.1$ cm, we have
\begin{equation}
  \mathrm{Re}=0.0123353,\quad  \mathrm{Pe}_{s}=8.22353,\quad  \mathrm{Fr}=0.000124605, \quad \mathrm{Sc}=1000, \quad \frac{\mathrm{Re}}{\mathrm{Fr^{2}}}=794468.
\end{equation}
For a larger channel width $L=1$ cm, we have
\begin{equation}
  \mathrm{Re}=0.123353, \quad  \mathrm{Pe}_{s}=82.2353, \quad \mathrm{Fr}=0.0000394036, \quad  \mathrm{Sc}=1000,  \quad \frac{\mathrm{Re}}{\mathrm{Fr^{2}}}=7.94468 \times 10^7.
\end{equation}
We can see that the Reynolds number is small, and the gravity term is important in the governing equation. 

\section{Flow equation}
\label{sec:Flow equation}

The Boussinesq approximation is commonly employed in the analysis of buoyancy-driven flow \cite{deen1998analysis}, as well as in previous studies of steady diffusion-driven flow \cite{phillips1970flows,wunsch1970oceanic}. This approximation is valid when the relative change in density is small, i.e., $\partial_{z} \rho / \rho \ll 1$, which holds true for the above given parameters where the value is 0.007. Therefore, adopting the Boussinesq approximation is a reasonable choice. The  Boussinesq approximation states that the density variation is only important in the buoyancy term,   

\begin{equation}\label{eq:NS nondimensional 2D diffusion driven flow Boussinesq}
  \begin{aligned}
&\rho_{0} \left( \frac{1}{ \mathrm{Sc}}   \partial_{t}  v_{1}+ \mathrm{Re} v_{1}\partial_{y_{1}} v_{1} + \mathrm{Re} v_{3}\partial_{y_{3}} v_{1}\right) =\Delta v_{1} - \partial_{y_{1}} p- \frac{\mathrm{Re}}{\mathrm{Fr^{2}}}\rho  \sin \theta,  \\
&\rho_{0} \left( \frac{1}{ \mathrm{Sc}}  \partial_{t}  v_{3}+ \mathrm{Re} v_{1}\partial_{y_{1}} v_{3} +\mathrm{Re}  v_{3}\partial_{y_{3}} v_{3}\right) = \Delta v_{3} - \partial_{y_{3}} p- \frac{\mathrm{Re}}{\mathrm{Fr^{2}}}\rho  \cos \theta,  \\
& \partial_{t}T+ \mathrm{Pe}_{p}  \mathbf{v} \cdot \nabla T= \Delta T,\\
&\frac{1}{ \kappa_{2}}\partial_{t}\rho+ \mathrm{Pe}_{s}  \mathbf{v} \cdot \nabla \rho= \Delta\rho,\quad \left. \rho \right|_{|\mathbf{y}|\rightarrow \infty }=\rho_0- \Gamma_0 (y_{1}\sin \theta + y_{3} \cos \theta).\\
\end{aligned}
\end{equation}

Notice that, in this setup,  the flow is invariant under the translation in $y_1$-direction. Hence, we can assume the velocity only depends on $y_3$. Then, the incompressibility $\partial_{y_{1}}v_{1}+ \partial_{y_{3}}v_{3}=0$ becomes $\partial_{y_{3}}v_{2}=0$ which implies $v_3=0$. To further simplify the equations, we introduce the density perturbation $f (y_{3},t)$ which satisfies
\begin{equation}\label{eq:density decomposition}
\begin{aligned}
\rho  =\rho_{0}+f (y_{3},t)- \Gamma_0 (y_{1}\sin \theta + y_{3} \cos \theta).
\end{aligned}
\end{equation}
We also write the pressure as $p=p_{0}+\tilde{p}$, where $p_0$ balances the background density and solves the equation
\begin{equation}
\begin{aligned}
&\partial_{y_{1}} p_{0} = - \frac{\mathrm{Re}}{\mathrm{Fr^{2}}} \sin \theta(\rho_{0}- \Gamma_0 (y_{1}\sin \theta + y_{3} \cos \theta)) ,  \\
&\partial_{y_{3}} p_{0} = -\frac{\mathrm{Re}}{\mathrm{Fr^{2}}}\cos \theta (\rho_{0}- \Gamma_0 (y_{1}\sin \theta + y_{3} \cos \theta)).  \\
\end{aligned}
\end{equation}
Since the right hand side of the above equation is curl-free, the solution $p_{0}$ exists. In fact, we have
\begin{equation}
\begin{aligned}
p_{0}=  -\frac{\mathrm{Re}}{\mathrm{Fr^{2}}} \left(\rho_{0}y_{3} \cos \theta +\rho_{0}y_{1} \sin \theta - \Gamma_0 y_{1}y_{3}\cos \theta \sin \theta - \frac{\Gamma_0}{2} y_{3}^{2}   \cos^{2}\theta - \frac{\Gamma_0}{2} y_{1}^{2} \sin^{2} \theta   \right).
\end{aligned}
\end{equation}
Now, equation \eqref{eq:NS nondimensional 2D diffusion driven flow Boussinesq} becomes
\begin{equation}
\begin{aligned}
&\frac{1}{\mathrm{Sc}}\rho_{0} \partial_{t}v_1= \partial_{y_{3}}^2v_1- \partial_{y_1}\tilde{p}- \frac{\mathrm{Re}}{\mathrm{Fr}^2}f \sin \theta , \quad \left. v_1 \right|_{y_{3}=0,1 }=0, \quad \left. v_1 \right|_{t=0 }=0,\\
&0=-\partial_{y_3}\tilde{p}- \frac{\mathrm{Re}}{\mathrm{Fr}^2}f \cos \theta,  \\
&\frac{1}{\kappa_{2}}\partial_{t}f-\partial_{y_{3}}^{2} f= \mathrm{Pe}_{s}\Gamma_0 v_1\sin \theta, \quad \left. \partial_{y_3}f \right|_{y_{3}=0,1 }=\Gamma_0 \cos \theta, \quad \left. f \right|_{t=0}=0. \\
\end{aligned}
\end{equation}
Obviously, $\tilde{p}$ can be a function of $y_3$ only. Due to the non-dimensionalization, $\rho_0=1$. We obtain the following equation for analyzing
\begin{equation}\label{eq:diffusion-driven flow parallel-plate channel domain simplified}
\begin{aligned}
&\frac{1}{\mathrm{Sc}} \partial_{t}v_1- \partial_{y_{3}}^2v_1=- \frac{\mathrm{Re}}{\mathrm{Fr}^2}f \sin \theta, \quad \left. v_1 \right|_{y_{3}=0,1 }=0, \quad \left. v_1 \right|_{t=0 }=0,  \\
&\frac{1}{\kappa_{2}}\partial_{t}f-\partial_{y_{3}}^{2} f= \mathrm{Pe}_{s}\Gamma_0 v_1\sin \theta, \quad \left. \partial_{y_3}f \right|_{y_{3}=0,1 }=\Gamma_0 \cos \theta, \quad \left. f \right|_{t=0}=0. \\
\end{aligned}
\end{equation}
We can decouple $f$ and $v_1$ by differentiating the equation and obtain the following equations
\begin{equation}
\begin{aligned}
&\left(\frac{1}{\kappa_{2}} \partial_{t}- \partial_{y_{3}}^2 \right) \left( \frac{1}{\mathrm{Sc}} \partial_{t}- \partial_{y_{3}}^2 \right)v_1=- \frac{\Gamma_0 \mathrm{Re} \mathrm{Pe}_{s} \left(  \sin \theta \right)^{2}}{\mathrm{Fr}^2}v_{1} , \quad \left. v_1 \right|_{y_{3}=0,1 }=0, \quad \left. v_1 \right|_{t=0 }=0,  \\
&\left( \frac{1}{\mathrm{Sc}}\partial_{t}- \partial_{y_{3}}^2 \right) \left( \frac{1}{\kappa_{2}} \partial_{t}- \partial_{y_{3}}^2 \right)f=- \frac{\Gamma_0 \mathrm{Re} \mathrm{Pe}_{s} \left(  \sin \theta \right)^{2}}{\mathrm{Fr}^2}f , \quad \left. \partial_{y_3}f \right|_{y_{3}=0,1 }=\Gamma_0 \cos \theta, \quad \left. f \right|_{t=0}=0. \\
\end{aligned}
\end{equation}

\begin{figure}
  \centering
    \subfigure[]{
    \includegraphics[width=0.46\linewidth]{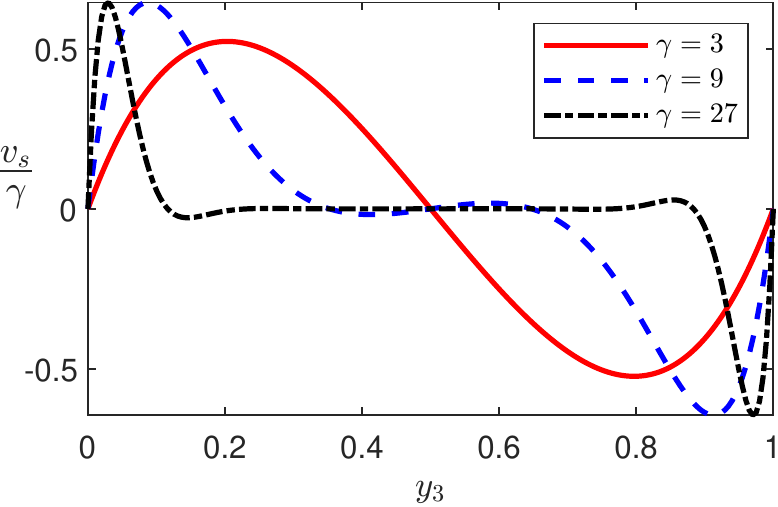}
  }
    \subfigure[]{
    \includegraphics[width=0.46\linewidth]{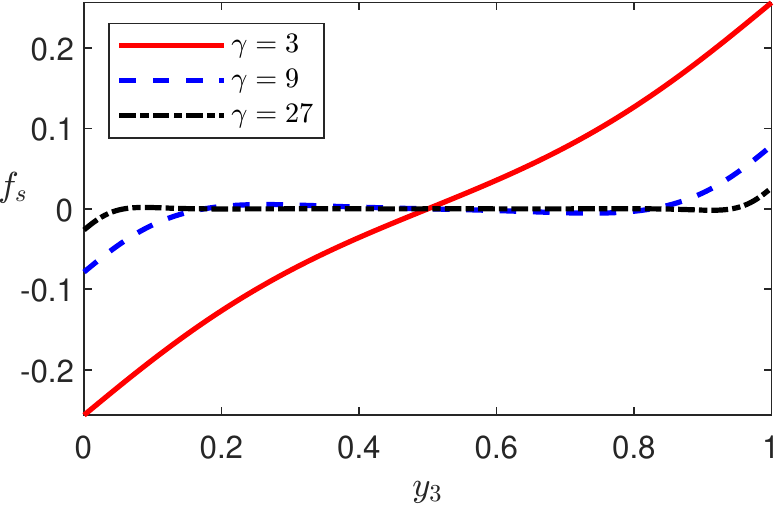}
  }

  \hfill
  \caption[]
  {(a) Normalized steady flow solution $\frac{v_{s}}{\gamma}$ for various parameter $\gamma$. (b) The perturbed density field $f_{s}$ for different $\gamma$.  }
  \label{fig:DiffusionDrivenFlowSteadySolution}
\end{figure}

To focus on the transient dynamics, we decompose the density perturbation and velocity into the steady part and the transient part, namely, $f= f_s+f_t$, $v_{1}=v_{s}+v_{t}$. We first consider the steady solution, which satisfies the following equation
\begin{equation}
\begin{aligned}
&\partial_{y_{3}}^4v_s=  -\frac{\mathrm{Re} \mathrm{Pe}_{s}}{\mathrm{Fr}^2} \left( \sin \theta \right)^{2} \Gamma_0 v_{s}, \quad \left. v_s \right|_{y_{3}=0,1 }=0, \quad \left. \partial_{y_{3}}^{3}v_s \right|_{y_{3}=0,1 }= \frac{\mathrm{Re}}{\mathrm{Fr}^2} \Gamma_0 \sin \theta \cos \theta,   \\
&\partial_{y_{3}}^{4} f_{s}=- \frac{\mathrm{Re} \mathrm{Pe}_{s}}{\mathrm{Fr}^2} \left( \sin \theta \right)^{2} \Gamma_0 f_{s}, \quad \left. \partial_{y_3}f_{s} \right|_{y_{3}=0,1 }=\Gamma_0 \cos \theta, \quad \left. \partial_{y_2}^{2}f_{s} \right|_{y_{3}=0,1 }=0, \\
\end{aligned}
\end{equation}
We can solve it easily and obtain the solution
\begin{equation}
\begin{aligned}
  &f_{s}=\frac{\Gamma_0 \cos \theta (\cos (\gamma(1- y_{3})) \cosh (\gamma y_{3})-\cos (\gamma y_{3}) \cosh (\gamma(1- y_{3})))}{\gamma (\sin (\gamma)+\sinh (\gamma))},\\
  &v_{s}= \frac{2 \gamma  \cot (\theta )}{\mathrm{Pe}_{s}} \frac{ \sin (\gamma  y_{3}) \sinh (\gamma (1-  y_{3}))-\sin (\gamma (1- y_{3})) \sinh (\gamma  y_{3}) }{ \sin (\gamma )+\sinh (\gamma )},
\end{aligned}
\end{equation}
where $\gamma= \frac{ 1}{\sqrt{2}}\left( \frac{\mathrm{Re} \mathrm{Pe}_{s} \left( \sin \theta \right)^{2} \Gamma_0 }{\mathrm{Fr}^2} \right)^{\frac{1}{4}}$, which is consistent with the steady solution presented in \cite{phillips1970flows,heitz2005optimizing}. $\gamma^{-1}$ indicates the thickness of the boundary layer. As shown in figure \ref{fig:DiffusionDrivenFlowSteadySolution}, both the flow and the perturbed density are confined in a narrow region near the boundary for a large $\gamma$. In addition, both $f$ and $v_1$ are odd functions with respect to $y_{3}=\frac{1}{2}$.

When the channel gap thickness approach to the infinity, the system should  asymptotically converge to the case with the inclined plane. Indeed, as $\gamma\rightarrow \infty$, we have 
\begin{equation}
\begin{aligned}
  &f_{s}=-\Gamma_0 \cos \theta\frac{e^{- \gamma  y} \cos (\gamma  y)}{\gamma },\\
  &v_{s}= \frac{2 \gamma  \cot (\theta )}{\mathrm{Pe}_{s}} \frac{e^{-\gamma  y} \sin (\gamma  y)}{2 \gamma ^3},
\end{aligned}
\end{equation}
where is consistent with the solution presented in \cite{phillips1970flows}.

 The transient part of the density perturbation $f_t$ satisfies the equation
\begin{equation}\label{eq:diffusion-driven flow parallel-plate channel domain simplified transient density}
\begin{aligned}
  &\left(\left( \frac{1}{\mathrm{Sc}}\partial_{t}- \partial_{y_{3}}^2 \right) \left(\frac{1}{\kappa_{2}}  \partial_{t}- \partial_{y_{3}}^2 \right)+\frac{\Gamma_0 \mathrm{Re} \mathrm{Pe}_{s} \left(  \sin \theta \right)^{2}}{\mathrm{Fr}^2}   \right)f_t=0,\\
  & \left. \partial_{y_2}f_t \right|_{y_{3}=0,1 }=0,\quad \left. f_t \right|_{t=0 }=- f_s.\\
\end{aligned}
\end{equation}
We need one more condition to determine the solution. From $ \left( \frac{1}{\kappa_{2}}\partial_{t}-\partial_{y_{3}}^{2} \right) f= \mathrm{Pe}_{s}\Gamma_0 v_1\sin \theta$, we have $\left. \left( \frac{1}{\kappa_{2}}\partial_{t}-\partial_{y_{3}}^{2} \right) f  \right|_{t=0 }=0$ which implies $\left.\frac{1}{\kappa_{2}} \partial_{t}f_t \right|_{t=0 } = \left. \partial_{y_{3}}^{2} f_t \right|_{t=0 }+ \partial_{y_{3}}^{2} f_s=0$.  To shorten the expression, we denote $\phi_{0}=1$, $\lambda_{0}=0$ and $\phi_n= \sqrt{2} \cos n \pi y$, $\varphi_{n}= \sqrt{2}\sin n \pi y$, $\lambda_n=n^{2}\pi^{2}$, $ n\geq 1$ as the eigenfunctions and eigenvalues of the Laplace operator in the cross section of the parallel-plate channel with no-flux boundary condition and pure absorbing boundary condition, respectively. To be more specific, $(\lambda_n-\Delta) \phi_n=0$, $\left.\partial_{y_{3}} \phi \right|_{y_{3}=0,1 }=0$ and $(\lambda_n-\Delta) \varphi_n=0$, $\left. \varphi \right|_{y_{3}=0,1 }=0$. Either $\left\{ \phi_n \right\}_{n=0}^{\infty}$ or $\left\{ \varphi_{n} \right\}_{n=1}^{\infty} \cup \left\{ 1 \right\}$  form an orthogonal basis on the  cross section $\Omega$ with respect to the inner product  $\left\langle f (y_{3}), g (y_{3}) \right\rangle= \int\limits_0^1 f (y_{3}) g(y_{3}) \mathrm{d} y_{3}$. For the velocity, we prefer to use $\varphi_{n}$, since the linear combination of them satisfies the boundary condition automatically. With the same argument, we prefer to use $\phi_{n}$ to represent the perturbed density field.    The straightforward calculation yields
\begin{equation}
\begin{aligned}
&  f_t= -\sum\limits_{n=1}^{\infty}  \left\langle f_s, \phi_{n} \right\rangle\phi_{n} (y_{3})e^{-\frac{1}{2} (\text{Sc}+\kappa_{2})  \lambda _n t}\left(\cosh \left(\frac{ a_{n} t}{2}\right)  +\frac{\lambda _n \left( \text{Sc}+\kappa_{2} \right)}{ a_{n}} \sinh \left(\frac{ a_{n}t}{2}  \right)\right),\\
&    a_{n}= \sqrt{(\text{Sc}-\kappa_{2})^2 \lambda _n^2-16\gamma ^4 \text{Sc}\kappa_{2}},\\
&\left\langle f_s, \phi_{n} \right\rangle= \frac{\sqrt{2}\Gamma_0 \cos \theta   \left((-1)^n-1\right) \left(\sin (\gamma ) \left(\pi ^2 n^2-2 \gamma ^2\right)+\sinh (\gamma ) \left(2 \gamma ^2+\pi ^2 n^2\right)\right)}{\left(4 \gamma ^4+\pi ^4 n^4\right) (\sin (\gamma )+\sinh (\gamma ))}.  
\end{aligned}
\end{equation}
Then the cosine expansion of $v_1$ is available from the relation \eqref{eq:diffusion-driven flow parallel-plate channel domain simplified}. We can obtain the sine expansion of the velocity  using the same strategy. The transient part of the velocity component in $y_{1}$ direction  satisfies the equation
\begin{equation}
\begin{aligned}
  &\left(\left( \frac{1}{\mathrm{Sc}}\partial_{t}- \partial_{y_{3}}^2 \right) \left( \frac{1}{\kappa_{2}} \partial_{t}- \partial_{y_{3}}^2 \right)+\frac{\Gamma_0 \mathrm{Re} \mathrm{Pe}_{s} \left(  \sin \theta \right)^{2}}{\mathrm{Fr}^2}   \right)v_t=0,\\
  & \left. v_t \right|_{y_{3}=0,1 }=0,\quad \left. v_t \right|_{t=0 }=- v_s.\\
\end{aligned}
\end{equation}
We need one more condition to determine the solution. Based on $ \left(\frac{1}{\mathrm{Sc}} \partial_{t}-\partial_{y_{3}}^{2} \right) v_{1}= -\frac{\mathrm{Pe_{2}}}{\mathrm{Fr}^{2}}f \sin \theta$, we have $\left. \left( \frac{1}{\mathrm{Sc}} \partial_{t}-\partial_{y_{3}}^{2} \right) v  \right|_{t=0 }=0$ which implies $\left. \frac{1}{\mathrm{Sc}} \partial_{t}v_{t} \right|_{t=0 } = \left. \partial_{y_{3}}^{2} v_t \right|_{t=0 }+ \partial_{y_{3}}^{2} v_s=0$. We have the series representation
\begin{equation}\label{eq:diffusion driven flow transient part series}
\begin{aligned}
&  v_t=  -\sum\limits_{n=1}^{\infty}  \left\langle v_s, \varphi_{n} \right\rangle\varphi_{n} (y_{3})e^{-\frac{1}{2} (\text{Sc}+\kappa_{2})  \lambda _n t}\left(\cosh \left(\frac{ a_{n} t}{2}\right)  +\frac{\lambda _n \left( \text{Sc}+\kappa_{2} \right)}{ a_{n}} \sinh \left(\frac{ a_{n}t}{2}  \right)\right),\\
&\left\langle v_{s},\varphi_{n} \right\rangle=-\frac{2 \gamma  \cot (\theta )}{\mathrm{Pe}_{s}} \frac{2 \sqrt{2} \pi  \gamma ^2 \left((-1)^n+1\right) n (\cos (\gamma )-\cosh (\gamma ))}{\left(4 \gamma ^4+\pi ^4 n^4\right) (\sin (\gamma )+\sinh (\gamma ))}.
\end{aligned}
\end{equation}

In a system with the inclined plane, the transient part of the diffusion-driven flow decays algebraically and exhibits oscillation behavior for all Schmidt numbers \cite{harabin2016diffusively}. Unlike the semi-infinite domain, here, the transient part of the flow vanishes exponentially. Moreover, $v_{t}$ can be a monotonic function for some parameters and oscillatory for other parameter combinations. For instance, in the limiting case $\mathrm{Sc}=\infty$, we have 
\begin{equation}
\begin{aligned}
&  f_t= -\sum\limits_{n=1}^{\infty} e^{-t \kappa_{2}\left( \frac{4\gamma^{4}}{ \lambda_{n}}+\lambda_{n}\right)} \phi_n (y_{3}) \left\langle f_s, \phi_{n} \right\rangle,\\
&    v_t= -\sum\limits_{n=1}^{\infty} e^{-t\kappa_{2} \left( \frac{4\gamma^{4}}{ \lambda_{n}}+\lambda_{n}\right)} \varphi_n (y_{3}) \left\langle v_s, \varphi_{n} \right\rangle.\\
\end{aligned}
\end{equation}
In this case,  $a_n$ is a real number for all $n$. Since $\left\langle v_{s},\varphi_{n} \right\rangle$ is positive definite, $v_{1}=v_{s}+v_{t}$ converges to the steady solution $v_{s}$ monotonically.

 When  $\mathrm{Sc}=\kappa_{2}$, we have a simpler expression
\begin{equation}
  \begin{aligned}
     f_t=& -\sum\limits_{n=1}^{\infty} \left\langle f_s, \phi_{n} \right\rangle\phi_{n} (y_{3})e^{-  \lambda _n \kappa_{2} t} \left( \cos \left( 2\gamma ^2 \kappa_{2} t \right)+ \frac{\lambda_{n}}{2\gamma^{2}} \sin \left( 2 \gamma^2 \kappa_{2} t   \right)\right),\\
v_t=& -\sum\limits_{n=1}^{\infty}  \left\langle v_s, \varphi_{n} \right\rangle\varphi_{n} (y_{3})e^{-  \lambda _n \kappa_{2} t}\left( \cos \left( 2\gamma ^2 \kappa_{2} t \right)+\frac{\lambda_{n}}{2\gamma^{2}} \sin \left( 2 \gamma^2 \kappa_{2} t   \right)\right).
\end{aligned}
\end{equation}
In this case, $a_n$ is  a pure imaginary number for all $n$ and the flow solution includes oscillatory terms. It is easy to show that the oscillation terms only appear if $a_{n}^{2}<0$ for some $n$, which can only happen when the parameters satisfy
\begin{equation}\label{eq:relationSc}
\begin{aligned}
\exists n\in \mathbb{Z}^{+},\frac{\kappa _2 \left(8 \gamma ^4-4 \sqrt{4 \gamma ^8+\pi ^4 \gamma ^4 n^4}+\pi ^4 n^4\right)}{\pi ^4 n^4}< \mathrm{Sc}< \frac{\kappa _2 \left(8 \gamma ^4+4 \sqrt{4 \gamma ^8+\pi ^4 \gamma ^4 n^4}+\pi ^4 n^4\right)}{\pi ^4 n^4},
\end{aligned}
\end{equation}
The inclined plane can be considered as a
tilted parallel-plate channel domain with the infinite channel width. As the channel width $L$ increases, $\mathrm{Pe}_2$ and $\gamma$ increases. For a large $\gamma$, we have the asymptotic expansion
\begin{equation}
\begin{aligned}
\frac{\kappa _2 \left(8 \gamma ^4-4 \sqrt{4 \gamma ^8+\pi ^4 \gamma ^4 n^4}+\pi ^4 n^4\right)}{\pi ^4 n^4}=&\frac{\pi ^4n^4  \kappa _2 }{16 \gamma ^4}+\mathcal{O}\left(\gamma^{-6}\right), \\
\frac{\kappa _2 \left(8 \gamma ^4+4 \sqrt{4 \gamma ^8+\pi ^4 \gamma ^4 n^4}+\pi ^4 n^4\right)}{\pi ^4 n^4}= &\frac{16 \gamma ^4 \kappa _2}{\pi ^4 n^4}+2 \kappa _2+\mathcal{O}\left(\gamma^{-1}\right). \\
\end{aligned}
\end{equation}
Therefore, in the large channel width limit, we observe the oscillation for all $\mathrm{Sc}$, which is consistent with the conclusions for the inclined plane problem \cite{harabin2016diffusively}.

\begin{figure}
  \centering
  \subfigure[$v_{t}$]{
    \includegraphics[width=0.46\linewidth]{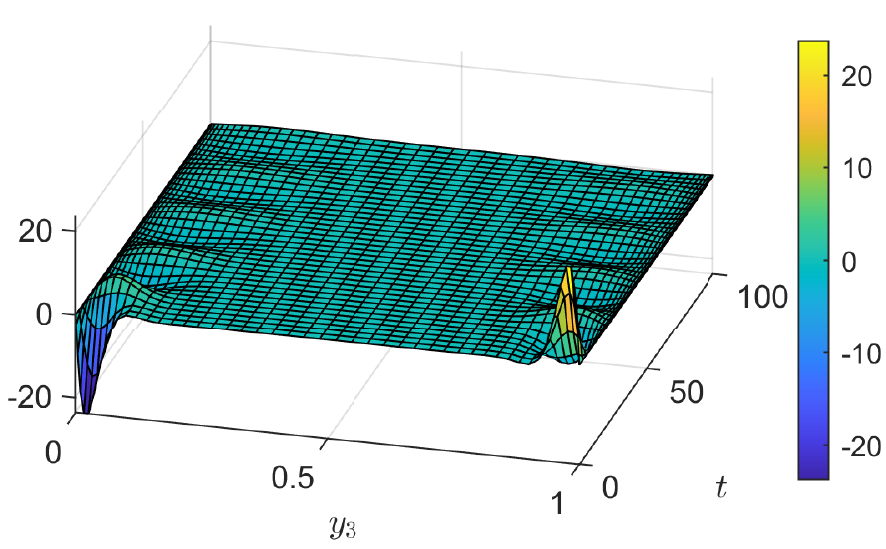}
  }
    \subfigure[$v_{1}=v_{s}+v_{t}$]{
    \includegraphics[width=0.46\linewidth]{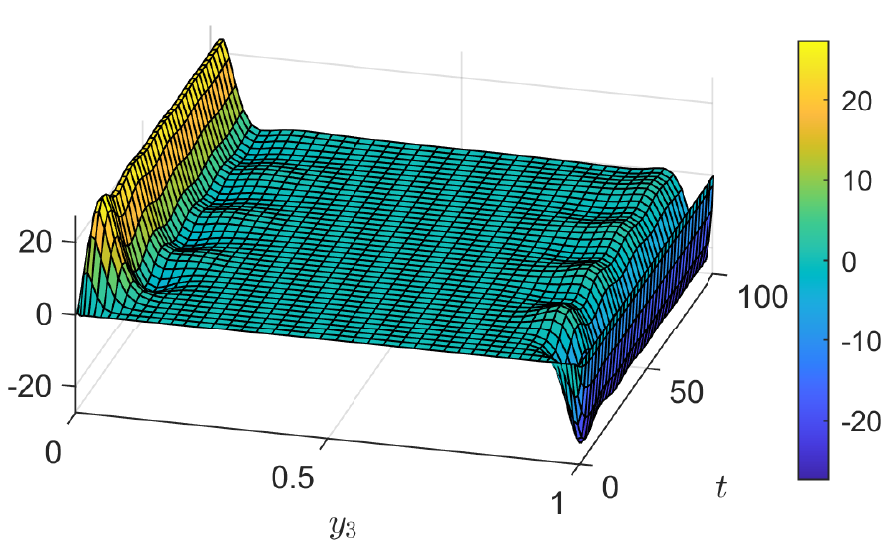}
  }
  \hfill
  \caption[]
  {Panel (a) The transient part of the diffusion-driven flow provided in equation \eqref{eq:diffusion driven flow transient part series}. We use the terms with $n\leq 200$ in the series. We verify the truncation error is small enough by doubling the number of terms. Panel (b) The unsteady diffusion-driven flow solution.  The parameters are $\mathrm{Sc}=\kappa_{2}=10^{-4}$, $\gamma=12\pi$, $\mathrm{Pe}_{s}=1$, $\theta= \frac{\pi}{4}$. }
  \label{fig:DiffusionDrivenFlowUnsteady}
\end{figure}

Next, we seek the parameters for observing pronounced oscillations in the time-dependent flow solution. The flow transient time scale (set by the longest lived mode) is $\frac{2}{(\text{Sc}+\kappa_{2})\pi^{2}}$.  The period of the associated oscillating term is $\frac{4\pi}{\sqrt{16\gamma ^4 \text{Sc}\kappa_{2}-(\text{Sc}-\kappa_{2})^2 \pi^4}}$.  We are interested in maximizing the number of oscillations in this time interval which can be done by maximizing the ratio of these two time scales $\frac{2(\text{Sc}+\kappa_{2})\pi^{3}}{\sqrt{16\gamma ^4 \text{Sc}\kappa_{2}-(\text{Sc}-\kappa_{2})^2 \pi^4}}$, which is the number of periods that we can observed within the transient time scale. In fact, when $\mathrm{Sc}=\kappa_{2}$, this quantity reaches its maximum value $\frac{\gamma ^2}{\pi ^3}$. Figure \ref{fig:DiffusionDrivenFlowUnsteady}  shows the evolution of the time-dependent diffusion-driven flow solution with $\mathrm{Sc}=\kappa_{2}$. The transient part of the flow $v_{t}$ is large near the boundary at a short time scale and then has oscillations with a relatively smaller amplitude. The oscillation amplitude is comparable to the magnitude of the steady solution. Therefore, from panel (b) of figure  \ref{fig:DiffusionDrivenFlowUnsteady}, we can see that the full flow solution has visible fluctuations.  We remark that small values of $\mathrm{Sc}$ and $\kappa_2$ are possible if the stratified scalar is the temperature and the passive scalar is the salt solute, since the thermal diffusivity for liquid metals are generally of the order of $1$ cm$^2$/s whereas the salt diffusivity is at the order of $10^{-5}$ cm$^{2}$/s.

Lastly, the original coupled equations for the velocity and perturbed density involve elliptic operators and first-order time derivatives, and at first glance, may appear similar to elliptic equations.  However,  the decoupled system \eqref{eq:diffusion-driven flow parallel-plate channel domain simplified transient density} reveals a hyperbolic equation with a second-order time derivative, leading to distinct properties compared with elliptic equations. To illustrate the different, we compare equation \eqref{eq:diffusion-driven flow parallel-plate channel domain simplified transient density} with the case of a second order diffusion problem with a Laplace-Beltrami operator using eigenfunction expansion with modes, $\phi_{n}(y)e^{-\lambda_{n}t }$.  According to the Sturm-Liouville theory, the eigenfunction expansion has temporally decaying modes indexed by the well-ordered eigenvalues of a one-dimensional Laplace-Beltrami operator, $\lambda_n<\lambda_{n+1}$. For each mode, $\lambda_n$, the associated eigenfunction, $\phi_n(y)$ has exactly $n-1$ zeros, (notice that the higher dimensional results are different \cite{berkolaiko2022stability}).  Interestingly,  the operator in equation \eqref{eq:diffusion-driven flow parallel-plate channel domain simplified transient density} doesn't have this property.  For example, when $\kappa_{2}=1$, $\gamma=3$, $\mathrm{Sc}=\frac{1}{10}$, the coefficients of $\varphi_{1} (y_{3}) =\sqrt{2}\cos(\pi y_{3})$ and $\varphi_{2} (y_{3}) =\sqrt{2} \cos(2 \pi y_{3})$ in equation \eqref{eq:diffusion driven flow transient part series} are, respectively,
\begin{equation}
  \begin{aligned}
    & \left\langle v_s, \varphi_{1} \right\rangle e^{-\frac{11 \pi ^2 t}{20} } \left(\frac{11 \pi ^2 \sin \left(\frac{1}{2} \sqrt{\frac{648}{5}-\frac{81 \pi ^4}{100}} t\right)}{10 \sqrt{\frac{648}{5}-\frac{81 \pi ^4}{100}}}+\cos \left(\frac{1}{2} \sqrt{\frac{648}{5}-\frac{81 \pi ^4}{100}} t\right)\right),\\
    & \left\langle v_s, \varphi_{2} \right\rangle e^{-\frac{11 \pi ^2 t}{5}} \left(\frac{22 \pi ^2 \sinh \left(\frac{1}{2} \sqrt{\frac{324 \pi ^4}{25}-\frac{648}{5}} t\right)}{5 \sqrt{\frac{324 \pi ^4}{25}-\frac{648}{5}}}+\cosh \left(\frac{1}{2} \sqrt{\frac{324 \pi ^4}{25}-\frac{648}{5}} t\right)\right)\\
    &\sim \left\langle v_s, \varphi_{2} \right\rangle \left(\frac{22 \pi ^2}{10 \sqrt{\frac{324 \pi ^4}{25}-\frac{648}{5}}}+\frac{1}{2}\right) e^{\left(\frac{1}{2} \sqrt{\frac{324 \pi ^4}{25}-\frac{648}{5}}-\frac{11 \pi ^2}{5}\right) t}\\
&\quad +\mathcal{O} \left(e^{-\left(\frac{1}{2} \sqrt{\frac{324 \pi ^4}{25}-\frac{648}{5}}+\frac{11 \pi ^2}{5}\right) t}\right), \quad t\rightarrow \infty.
\end{aligned}
\end{equation}
Since $-\frac{11 \pi ^2}{20}  \approx -5.42828$ and $\frac{1}{2} \sqrt{\frac{324 \pi ^4}{25}-\frac{648}{5}}-\frac{11 \pi ^2}{5} \approx -4.88442$, the coefficient of $\varphi_2$ decays slower than the coefficient of $\varphi_{1}$ at long times, but has more spatial oscillations.

\subsection{Quasi Boussinesq approximation}
\begin{figure}
  \centering
  \subfigure[]{
    \includegraphics[width=0.46\linewidth]{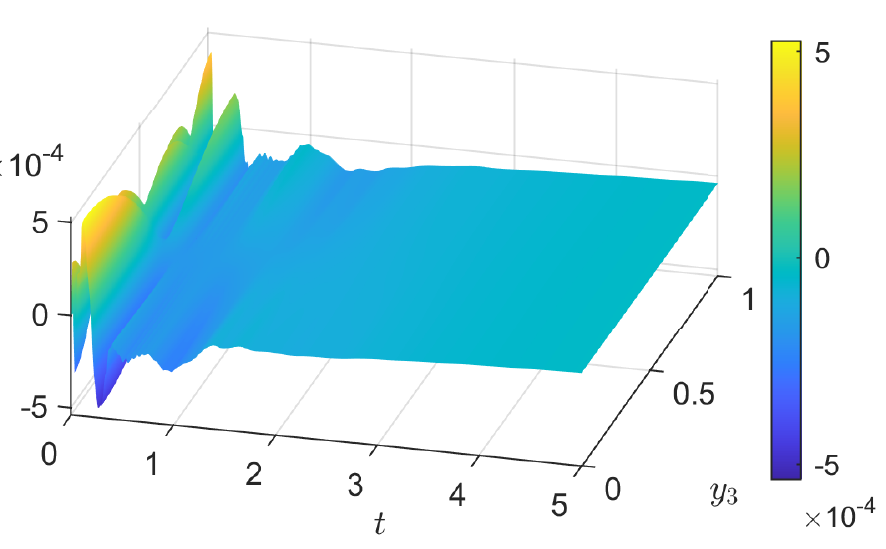}
  }
    \subfigure[]{
    \includegraphics[width=0.46\linewidth]{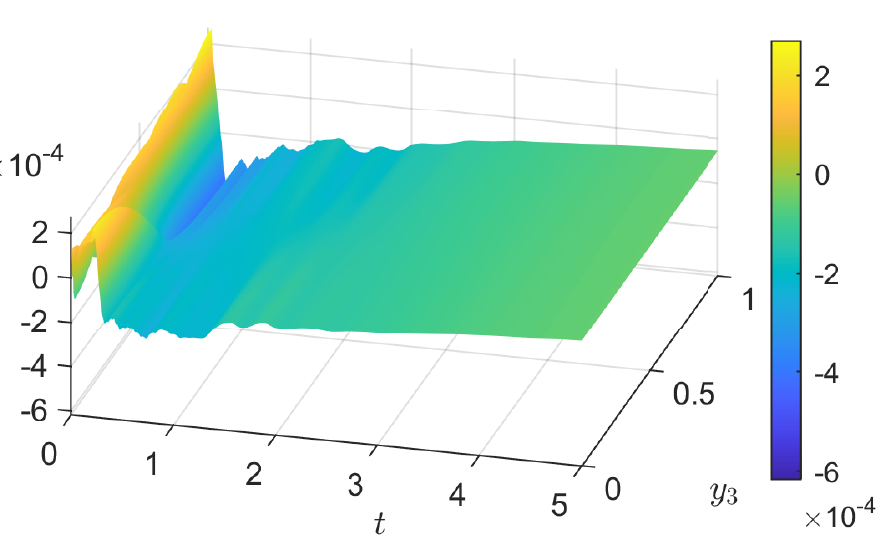}
  }
  \hfill
  \caption[]
  {The relative difference between the solution of equation \eqref{eq:diffusion-driven flow parallel-plate channel domain simplified} and the solution of equation \eqref{eq:diffusion-driven flow parallel-plate channel domain simplified decoupled  quasi} with $\theta=\frac{\pi}{4}$ in panel (a) and $\theta= \frac{\pi}{40}$ in panel (b). The rest parameters are $\gamma=1$, $\Gamma_{0}=1$, $\kappa_{2}=1$, $\mathrm{Sc}=1$. The relative difference is defined as $ \frac{f_{1}-f_{2}}{max_{y_{3},t} (f_{1})}$, where $f_{1}$ denotes the solution of equation \eqref{eq:diffusion-driven flow parallel-plate channel domain simplified}  and $f_{2}$ denotes the solution of equation \eqref{eq:diffusion-driven flow parallel-plate channel domain simplified decoupled  quasi}
  }
  \label{fig:DiffusionDrivenFlowQuasiErr}
\end{figure}

In the preceding section, we demonstrated how the Boussinesq approximation can simplify the problem and capture the nontrivial dynamics of the system, enabling us to obtain an exact solution for the unsteady shear flow. This approximation assumes a constant density function, denoted by $\rho_0$, in the time derivative term in equation \eqref{eq:NS nondimensional 3D diffusion driven flow rotated coordinate}. To further improve our understanding and capture more comprehensive behavior, we introduce an alternative approximation in this subsection that accounts for density variations in the $y_3$ direction, namely,
\begin{equation}
  \begin{aligned}
&\left( \rho_{0}+f (y_{3},t)- \Gamma_0  y_{3} \cos \theta \right) \left( \frac{1}{ \mathrm{Sc}}   \partial_{t}  v_{1}+ \mathrm{Re} v_{1}\partial_{y_{1}} v_{1} + \mathrm{Re} v_{3}\partial_{y_{3}} v_{1}\right) =\Delta v_{1} - \partial_{y_{1}} p- \frac{\mathrm{Re}}{\mathrm{Fr^{2}}}\rho  \sin \theta,  \\
&\left( \rho_{0}+f (y_{3},t)- \Gamma_0  y_{3} \cos \theta \right) \left( \frac{1}{ \mathrm{Sc}}  \partial_{t}  v_{3}+ \mathrm{Re} v_{1}\partial_{y_{1}} v_{3} +\mathrm{Re}  v_{3}\partial_{y_{3}} v_{3}\right) = \Delta v_{3} - \partial_{y_{3}} p- \frac{\mathrm{Re}}{\mathrm{Fr^{2}}}\rho  \cos \theta,  \\
& \partial_{t}T+ \mathrm{Pe}_{p}  \mathbf{v} \cdot \nabla T= \Delta T,\\
&\frac{1}{ \kappa_{2}}\partial_{t}\rho+ \mathrm{Pe}_{s}  \mathbf{v} \cdot \nabla \rho= \Delta\rho,\quad \left. \rho \right|_{|\mathbf{y}|\rightarrow \infty }=\rho_0- \Gamma_0 (y_{1}\sin \theta + y_{3} \cos \theta).\\
\end{aligned}
\end{equation}
This is a valid approximation when $\theta \ll 1$, as $\rho \approx \rho_{0}+f (y_{3},t)- \Gamma_0  y_{3} \cos \theta$.  This approximation retains the most advantages of the Boussinesq approximation in analysis. First, we can still find the solution that only depends on $y_{3}$, resulting the following equation:
\begin{equation}\label{eq:diffusion-driven flow parallel-plate channel domain simplified quasi}
\begin{aligned}
&\frac{1}{\mathrm{Sc}} \left( 1+f - \Gamma_0  y_{3} \cos \theta \right) \partial_{t}v_1- \partial_{y_{3}}^2v_1=- \frac{\mathrm{Re}}{\mathrm{Fr}^2}f \sin \theta, \quad \left. v_1 \right|_{y_{3}=0,1 }=0, \quad \left. v_1 \right|_{t=0 }=0,  \\
&\frac{1}{\kappa_{2}}\partial_{t}f-\partial_{y_{3}}^{2} f= \mathrm{Pe}_{s}\Gamma_0 v_1\sin \theta, \quad \left. \partial_{y_3}f \right|_{y_{3}=0,1 }=\Gamma_0 \cos \theta, \quad \left. f \right|_{t=0}=0. \\
\end{aligned}
\end{equation}
Here, $\rho_0$ is set to 1 due to non-dimensionalization. Second, we can also decouple $f$ and $v_1$ by differentiating the equation and obtain the following equation
\begin{equation}\label{eq:diffusion-driven flow parallel-plate channel domain simplified decoupled quasi}
\begin{aligned}
  &\left( \frac{\frac{1}{\mathrm{Sc}} \left( 1+f - \Gamma_0  y_{3} \cos \theta \right)}{\mathrm{Sc}}\partial_{t}- \partial_{y_{3}}^2 \right) \left( \frac{1}{\kappa_{2}} \partial_{t}- \partial_{y_{3}}^2 \right)f=- 4\gamma^{4}f,\\
  & \left. \partial_{y_3}f \right|_{y_{3}=0,1 }=\Gamma_0 \cos \theta, \quad \left. f \right|_{t=0}=0. \\
\end{aligned}
\end{equation}
Once we have obtained the perturbed density field, we can use it to compute the velocity field with equation \eqref{eq:diffusion-driven flow parallel-plate channel domain simplified quasi}. The steady-state solutions of equations \eqref{eq:diffusion-driven flow parallel-plate channel domain simplified} under the Boussinesq approximation is the same as the solution of equation \eqref{eq:diffusion-driven flow parallel-plate channel domain simplified quasi}, but their transient dynamics differ. Due to the nonlinearity of the problem, it is difficult to find an exact analytical solution of equation  \eqref{eq:diffusion-driven flow parallel-plate channel domain simplified quasi}, and here we numerically compute the solutions using \texttt{NDSolve} in Mathematica. We plot the relative difference between the solutions obtained from equations \eqref{eq:diffusion-driven flow parallel-plate channel domain simplified} and \eqref{eq:diffusion-driven flow parallel-plate channel domain simplified quasi} in Figure \ref{fig:DiffusionDrivenFlowQuasiErr}. For both large and small inclination angles, the relative differences of the perturbed density field is around $10^{-4}$, demonstrating that the system dynamics are not significantly affected by the transverse density variation in the time derivative term of the governing equation in this parameter regimes.  This demonstrates the validity of the Boussinesq approximation for small angles.

\section{Dispersion induced by the unsteady diffusion-driven flow}
\label{sec:Effective diffusivity}

In this section, we focus on the evolution of passive scalar under the advection of the unsteady diffusion-driven flow. The well-known Taylor dispersion \cite{taylor1953dispersion,aris1956dispersion} shows that as the flow acts to smear out the concentration distribution in the direction of the flow, it enhances the dispersion rate of the concentration distribution at which it spreads in that direction. Additionally, many approaches demonstrated that the distribution of a diffusing passive tracer under the shear flow advection is approximately governed by a diffusion equation with an effective diffusion coefficient at long-times, such as Hermite polynomial series expansion \cite{chatwin1970approach}, homogenization theory \cite{wu2014approach,camassa2010exact}, Aris moment approach \cite{aris1956dispersion,vedel2014time,vedel2012transient,ding2021enhanced}, center manifold theory \cite{mercer1990centre,wang2013self,ding2022determinism} and the moment reconstruction \cite{ding2021ergodicity,camassa2021persisting}.

We first formulate the approximation theory of the Taylor dispersion. The reader can find more details in \cite{ding2022determinism}. The effective equation for the governing equation of passive scalar \eqref{eq:NS nondimensional 3D diffusion driven flow} at long times is
\begin{equation}
\begin{aligned}
\partial_{t}T+ \mathrm{Pe}_{p}\bar{v}_{1}\partial_{y_{1}}T= \kappa_{\mathrm{eff}} (t)\partial_{y_{1}}^{2}T,\quad \kappa_{\mathrm{eff}}=1+ \mathrm{Pe}_{p} \left\langle v_{1} T_{1} \right\rangle, \\
\end{aligned}
\end{equation}
 where $T_1$ is the solution of the auxiliary problem
\begin{equation}\label{eq:Aris moment 1 diffusion-driven flow}
\partial_{t} T_1- \partial_{y_{3}}^{2} T_{1}=\mathrm{Pe}_{p}(v_{1}-\bar{v}_{1}) , \quad T_1(y_{3},0)= 0, \quad \left. \partial_{y_{3}} T_{1} \right|_{ y_{3}=0,1}= 0.
\end{equation}
 If the initial condition of the passive scalar is a Gaussian function $\left. T \right|_{t=0 } = \frac{1}{\sqrt{2\pi \sigma}} e^{-\frac{y_{1}^{2}}{2 \sigma }}$,  then we have the exact formula for the variance 
\begin{equation}\label{eq:variance}
  \mathrm{Var} (\bar{T}) (t) = \mathrm{Var} (\bar{T}) (0) + 2\int\limits_0^{t} \kappa_{\mathrm{eff}} (s)\mathrm{d}s. 
\end{equation}
For general initial conditions, we have more exponential decaying terms in the variance formula. Equation \ref{eq:variance} is a valid approximation at long times. The exact variance formula can be found in \cite{vedel2014time,vedel2012transient}.

 Using the relation between the flow and density perturbation \eqref{eq:diffusion-driven flow parallel-plate channel domain simplified}, we have
\begin{equation}
\begin{aligned}
  \partial_{t} T_1- \partial_{y_{3}}^{2} T_{1}=\mathrm{Pe}_{p} \left( \frac{\frac{1}{\kappa_{2}}\partial_{t}f_{t}- \partial_{y_{3}}^{2} f_{t} }{\mathrm{Pe}_{s}\Gamma_{0} \sin \theta}+v_{s} \right) , \quad T_1(y_{3},0)= 0, \quad \left. \partial_{y_{3}} T_{1} \right|_{ y_{3}=0,1}= 0.
\end{aligned}
\end{equation}
The solution is
\begin{equation}
\begin{aligned}
  &T_{1}=\mathrm{Pe}_{p} \sum\limits_{n=1}^{\infty} \left(\frac{\left\langle f_{s},\phi_{n} \right\rangle b_{n}}{\mathrm{Pe}_{s}\Gamma_{0} \sin \theta} + \left\langle v_{s},\phi_{n} \right\rangle \left( \frac{1-e^{-\lambda_{nt}}}{\lambda_{n}} \right) \right) \phi_{n},\\
\end{aligned}
\end{equation}
where
\begin{equation}
\begin{aligned}
  & \left\langle v_{s},\phi_{n} \right\rangle = -\frac{2 \gamma  \cot (\theta )}{\mathrm{Pe}_{s}} \frac{\sqrt{2} \gamma  \left((-1)^n-1\right) \left(\sin (\gamma ) \left(2 \gamma ^2+\pi ^2 n^2\right)+\sinh (\gamma ) \left(2 \gamma ^2-\pi ^2 n^2\right)\right)}{\left(4 \gamma ^4+\pi ^4 n^4\right) (\sin (\gamma )+\sinh (\gamma ))},\\
 & b_{n}=\frac{-1}{\kappa _2 a_n \left(\lambda _n^2 \left(\kappa _2+\text{Sc}-2\right){}^2-a_n^2\right)}\left(e^{-t \lambda _n} \left(a_n^3-a_n \lambda _n^2 \left(\text{Sc}^2+\kappa _2 \left(-3 \kappa _2-2 \text{Sc}+4\right)\right)\right)+\right.\\
  &e^{-\frac{\kappa _2+\text{Sc}}{2}  \lambda _n  t}\left(a_n^3 \left(-\cosh \left(\frac{t a_n}{2}\right)\right)-a_n^2 \lambda _n \left(3 \kappa _2+\text{Sc}-2\right) \sinh \left(\frac{t a_n}{2}\right)\right.\\
  &+a_n \lambda _n^2 \left(\text{Sc}^2+\kappa _2 \left(-3 \kappa _2-2 \text{Sc}+4\right)\right) \cosh \left(\frac{t a_n}{2}\right)\\
  &\left.\left.+\lambda _n^3 \left(\text{Sc}-\kappa _2\right) \left(\kappa _2+\text{Sc}-2\right) \left(\kappa _2+\text{Sc}\right) \sinh \left(\frac{t a_n}{2}\right)\right)\right).\\
\end{aligned}
\end{equation}

Then, we have the series representation of the effective diffusion coefficient
\begin{equation}\label{eq:Time depedent effective diffusivity diffusion driven flow parallel-plate channel}
\begin{aligned}
&  \kappa_{\mathrm{eff}}=1+\mathrm{Pe}_{p}^{2}\sum\limits_{n=1}^{\infty} \left(\frac{\left\langle f_{s},\phi_{n} \right\rangle}{\mathrm{Pe}_{s}\Gamma_{0} \sin \theta}  b_{n}+ \left\langle v_{s},\phi_{n} \right\rangle \left( \frac{1-e^{-\lambda_{n}t}}{\lambda_{n}} \right) \right) \\
& \times  \left(\left\langle v_{s},\phi_{n} \right\rangle - \frac{\left\langle f_{s},\phi_{n} \right\rangle e^{-\frac{\kappa _2+\text{Sc}}{2}  \lambda _n t}}{\mathrm{Pe}_{s}\Gamma_{0} \sin \theta}  \left( \sinh \left(\frac{t a_n}{2}\right) \frac{  a_n^2+\lambda _n^2 \left(\kappa _2^2-\text{Sc}^2\right)}{2 \kappa _2 a_n}+\lambda _n \cosh \left(\frac{t a_n}{2}\right) \right)\right).\\
\end{aligned}
\end{equation}

To understand the contribution from the transient part of the flow solution, we compare it with the effective diffusion coefficient induced by the steady flow solution,
 \begin{equation}\label{eq:Steady part Time depedent effective diffusivity diffusion driven flow parallel-plate channel}
\begin{aligned}
\kappa_{\mathrm{eff},s}=1+\mathrm{Pe}_{p}^{2}\sum\limits_{n=1}^{\infty}  \left\langle v_{s},\phi_{n} \right\rangle   ^{2}\left( \frac{1-e^{-\lambda_{n}t}}{\lambda_{n}} \right),
\end{aligned}
\end{equation}
and the long time limit of the effective diffusion coefficient 
\begin{equation}\label{eq:effective diffusivity diffusion driven flow parallel-plate channel}
\begin{aligned}
  \kappa_{\mathrm{eff}} (\infty) &=1+  \frac{\mathrm{Pe}_{p}^{2} \cot ^2(\theta ) }{2 \gamma  \mathrm{Pe}^2_{2}(\sin (\gamma )+\sinh (\gamma ))^2}\left( \frac{5}{2} \sin (2 \gamma )+6 \gamma  \sin (\gamma ) \sinh (\gamma )+ \right.\\
&\left.   5 \cos (\gamma ) \sinh (\gamma )+\gamma  (\cosh (2 \gamma )-\cos (2 \gamma ))-5 \cosh (\gamma ) (\sin (\gamma )+\sinh (\gamma )) \right).
\end{aligned}
\end{equation}
As an example, in a realizable experiment of sodium fluorescein diffusing in stratified sodium chloride solution, the parameters could be  $g=980$  cm/s$^{2}$, $\mu=0.01$ gram/(cm.s), $\Gamma=0.007$ gram/cm$^{4}$, $\rho=1$ gram/cm$^{3}$, $\theta= \frac{\pi}{4}$. The diffusivity of sodium fluorescein  is $\kappa_{2}= 4.2\times10^{-6}$ cm$^{2}$/s  \cite{casalini2011diffusion}, and the diffusivity of sodium chloride  is $\kappa_{2}= 1.5\times 10^{-5}$ cm$^{2}$/s \cite{vitagliano1956diffusion}. Based on the formula of the effective diffusivity, we have $\kappa_{\mathrm{eff}} (\infty) =6.958$ for $L=0.1$ cm, $\kappa_{\mathrm{eff}} (\infty) =13.103$ for $L=1$ cm.  The difference between the diffusivities of passive scalar and stratified scalar could be much larger in temperature stratified experiments, where the enhanced effective diffusivity will be more significant.

\begin{figure}
  \centering
  \subfigure[]{
    \includegraphics[width=0.46\linewidth]{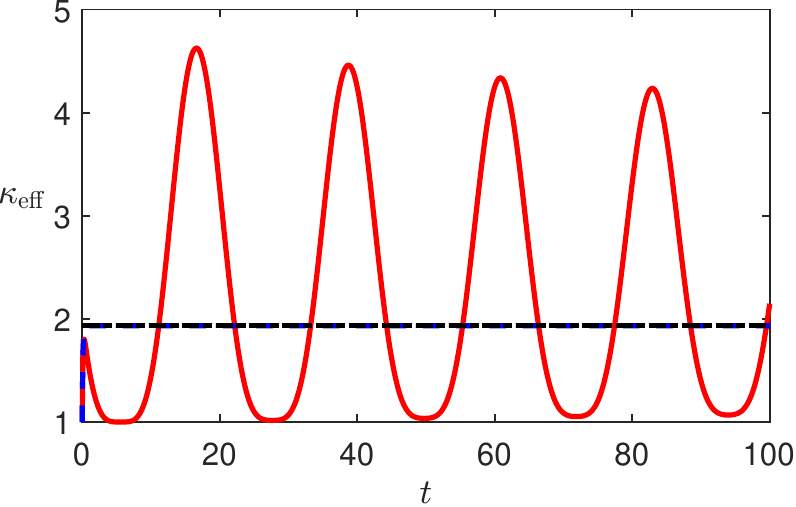}
  }
   \subfigure[]{
    \includegraphics[width=0.46\linewidth]{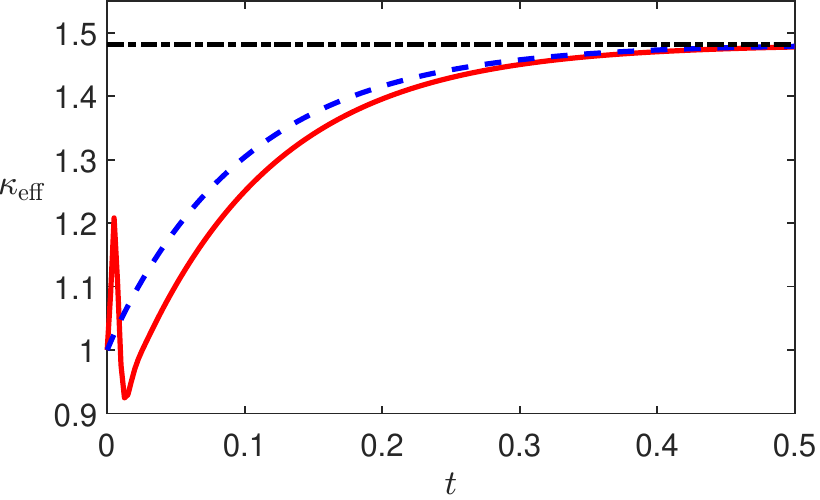}
  }
  \hfill
  \caption[]
  {Effective diffusion coefficient for various parameters. The red solid line indicates the time dependent effective diffusion coefficient induced by the unsteady diffusion driven flow. The formula is provided in equation \eqref{eq:Time depedent effective diffusivity diffusion driven flow parallel-plate channel}. The blue dashed line is the  effective diffusion coefficient contributed by the steady part of the flow which is calculated by equation \eqref{eq:Steady part Time depedent effective diffusivity diffusion driven flow parallel-plate channel}. The black dash-dot curve is the long time limit of the effective diffusion coefficient. The formula is provided in equation \eqref{eq:effective diffusivity diffusion driven flow parallel-plate channel}. Panel (a): The parameters are $\mathrm{Sc}=\kappa_{2}=10^{-4}$, $\gamma=12\pi$, $\mathrm{Pe}_{p}=\mathrm{Pe}_{s}=1$, $\theta= \frac{\pi}{4}$. The corresponding velocity field is presented in figure \ref{fig:DiffusionDrivenFlowUnsteady}. Panel (b):  The parameters are $\mathrm{Sc}=50$, $\kappa_{2}=1$, $\gamma=5$, $\mathrm{Pe}_{p}=\mathrm{Pe}_{s}=1$, $\theta= \frac{\pi}{4}$.   }
  \label{fig:DiffusionDrivenFlowUnsteadyKeff}
\end{figure}

Panel (a) in figure \ref{fig:DiffusionDrivenFlowUnsteadyKeff} shows the effective diffusion coefficient induced by the unsteady flow present in figure \ref{fig:DiffusionDrivenFlowUnsteady}, where the passive scalar molecular diffusivity is much smaller than the stratified scalar diffusivity. We can see that the effective diffusion coefficient induced by the steady flow converges to the limiting value at the passive scalar diffusion time scale $t=1$, while the effective diffusion coefficient induced by the unsteady diffusion-driven flow persists huge oscillations with the amplitude that is around twice of the limiting value at relatively larger time scales. Panel (b) in figure \ref{fig:DiffusionDrivenFlowUnsteadyKeff} shows the effective diffusion coefficient when the molecular diffusivity of passive scalar and stratified scalar are same. In this case, the effective diffusion coefficients induced by the steady and unsteady flow solution are closer. Interestingly, instead of enhancing the effective diffusion coefficient, the unsteady flow solution temporally reduces the effective diffusion coefficient below 1. In contrast, the steady flow creates dispersion enhancement for all parameters, which can be easily verified from equation \eqref{eq:Steady part Time depedent effective diffusivity diffusion driven flow parallel-plate channel}. Additionally, we emphasize this dispersion reducing phenomenon is not observed in the scalar transport with single-frequency time-varying periodic shear flows \cite{vedel2014time,vedel2012transient,ding2021enhanced}. We think this reduction is due to the interaction of different modes in the space-time decomposition of the shear flow. In the appendix, we present a simple shear flow example that consists of two modes and can reduce the dynamic effective diffusion coefficient below 1 at the earlier stages of the evolution.

\begin{figure}
  \centering
    \includegraphics[width=1\linewidth]{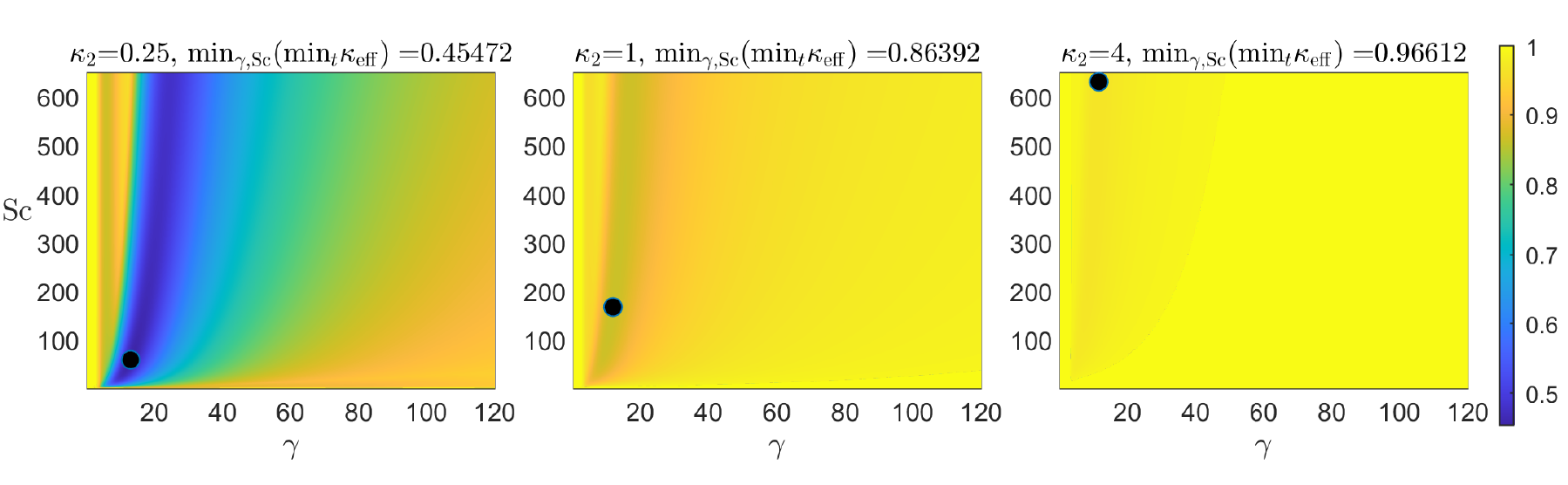}
    \includegraphics[width=1\linewidth]{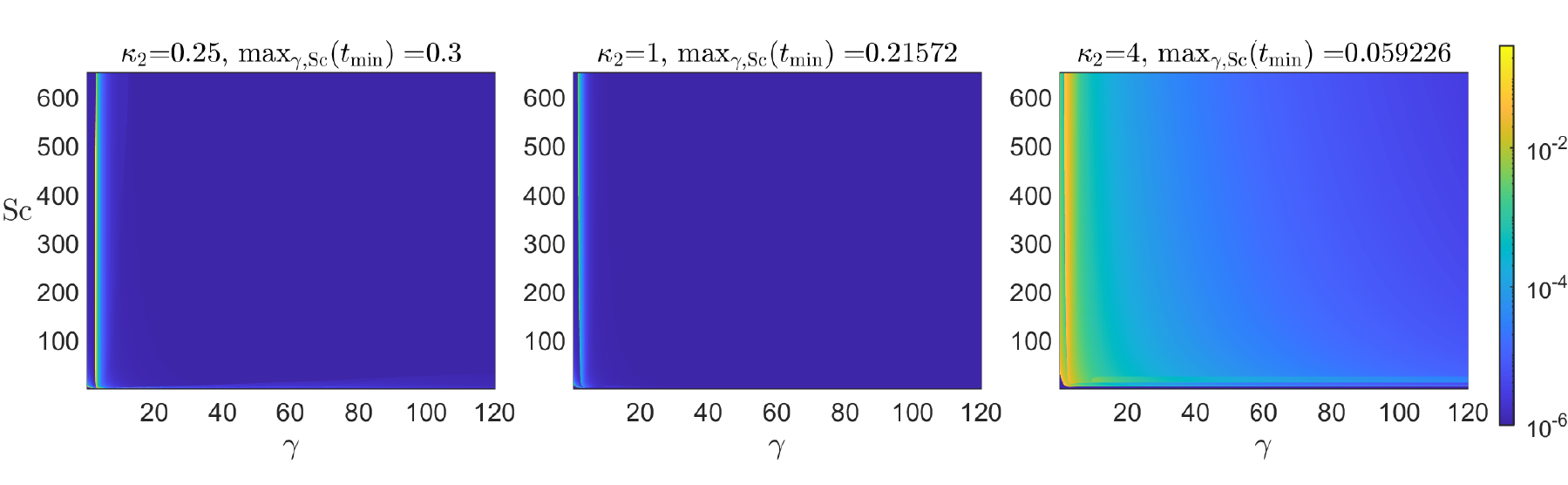}
  \caption[]
  {The first row shows the minimum value of the effective diffusion coefficient $\min_{t} \kappa_{\mathrm{eff}} (t)$  for $(\gamma, \mathrm{Sc})\in (0,120]\times (0,400]$, three different $\kappa_{2}$, $\theta=\frac{\pi}{4}$ and $\mathrm{Pe}_{p}= \mathrm{Pe}_{s}=1$.  The black dot indicates the location of the minimum value $\min_{\gamma, \mathrm{Sc}} \left( \min_{t} \kappa_{\mathrm{eff}}  \right)$ in this parameter regime. The optimal parameters for reaching $\min_{\gamma, \mathrm{Sc}} \left( \min_{t} \kappa_{\mathrm{eff}}  \right)$ are $\gamma =12.9407$ and $ \mathrm{Sc}= 58.5349$ for $\kappa_{2}=0.25$, $\gamma=11.6999$ and $\mathrm{Sc}=166.6348$ for $\kappa_{2}=1$ , $\gamma=10.16$ and $\mathrm{Sc}=400$ for $\kappa_{2}=4$.     The second row shows  $t_{\mathrm{min}}$ for $\kappa_{\mathrm{eff}}$ reaching the minimum value. We use the terms with $n\leq 259$ in the series.   }
  \label{fig:DiffusionDrivenFlowUnsteadyKeffSmallest}
\end{figure}

To further understand this phenomenon, we are interested in the dependence of the minimum effective diffusion coefficient $\min_{t} \kappa_{\mathrm{eff}} (t;\mathrm{Sc},\gamma)$ and the time for reaching its minimum value $t_{\mathrm{min}}=\mathrm{argmin}_{t}\kappa_{\mathrm{eff}} (t;\mathrm{Sc},\gamma)$ on $\mathrm{Sc}$ and $\gamma$.  We numerically search the minimum value and the results are summarized in figure \ref{fig:DiffusionDrivenFlowUnsteadyKeffSmallest}. We have several observations. First, in this parameter regime, as $\kappa_2$ decreases, $\min_{\gamma, \mathrm{Sc}} \left( \min_{t} \kappa_{\mathrm{eff}}  \right)$  decreases and $t_{\mathrm{min}}$ increases, which implies the dispersion reducing phenomenon is more significant for small $\kappa_{2}$, namely, when the passive scalar diffusivity is larger than the stratified scalar diffusivity.  Second,  $\min_{t} \kappa_{\mathrm{eff}} (t;\mathrm{Sc},\gamma)$ is considerably less than 1 for moderate $\gamma$ ($10 \sim 20$) and is closer to 1 for large $\gamma$.

\begin{figure}
  \centering
    \includegraphics[width=0.46\linewidth]{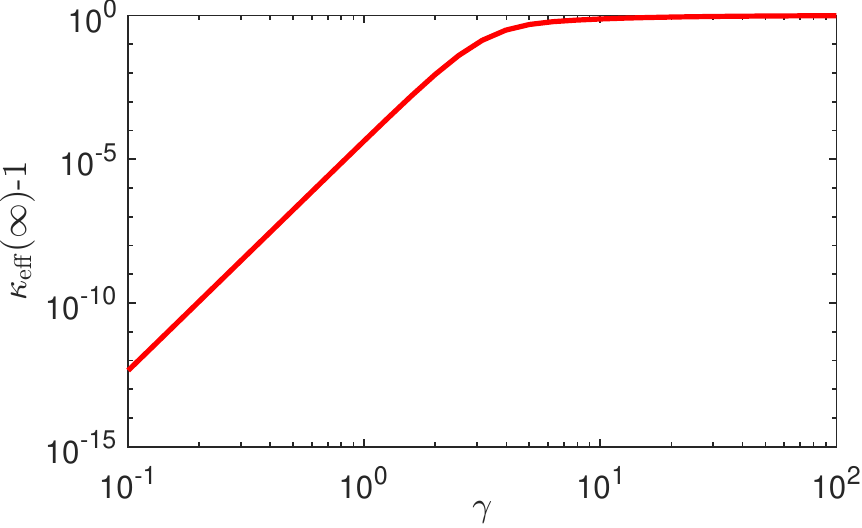}
  \hfill
  \caption[Enhanced effective diffusivity against the nondimensional parameter $\gamma$ ]
  {$\kappa_{\mathrm{eff}}-1$ against the nondimensional parameter $\gamma= \frac{ 1}{\sqrt{2}}\left( \frac{\mathrm{Re} \mathrm{Pe}_{s} \left( \sin \theta \right)^{2} \Gamma_0 }{\mathrm{Fr}^2} \right)^{\frac{1}{4}}$. The parameters are $\mathrm{Pe}_{p}=\mathrm{Pe}_{s}=1$ and $\theta= \frac{\pi}{4}$. }
  \label{fig:diffusionDrivenKeGamma}
\end{figure}

Next, we focus on the dispersion enhancement at long times. First, we consider the dependence of the enhancement on the parameter $\gamma$. We have the asymptotic expansion of the effective diffusion coefficient for large and small $\gamma$, 
\begin{equation}\label{eq:effective diffusion coefficient gamma limit}
\begin{aligned}
  \kappa_{\mathrm{eff}} (\infty) = &1+\frac{\mathrm{Pe}_{p}^{2}\cot ^2(\theta )}{\mathrm{Pe}_{s}^{2}} \left(1-\frac{5}{2 \gamma }+ \mathcal{O} (e^{-\gamma}) \right),\quad \gamma \rightarrow \infty,\\
\kappa_{\mathrm{eff}} (\infty) = &1+ \frac{\mathrm{Pe}_{p}^{2}\cot ^2(\theta )}{\mathrm{Pe}_{s}^{2}}\left(\frac{\gamma ^8}{22680}-\frac{2879 \gamma ^{12}}{4086482400}+O\left(\gamma ^{13}\right)  \right), \quad \gamma\rightarrow 0.  
\end{aligned}
\end{equation}
These asymptotic expansions suggest that the effective diffusion coefficient is bounded by $\frac{\mathrm{Pe}_{p}^{2}\cot ^2(\theta )}{\mathrm{Pe}_{s}^{2}} = \left(\frac{\kappa_{2}\cot(\theta )}{\kappa_{p}}  \right)^{2}$. In fact, $\gamma\rightarrow 0$ as the channel width $L$ vanishes and $\gamma\rightarrow \infty$ as $L \rightarrow \infty$. When the channel width is small, the diffusion-driven flow is too weak to enhance the scalar dispersion. When the channel width is large, the diffusion-driven flow is confined in the region near the boundary and is not efficient to transport the scalar located far away from the boundary.   Figure \ref{fig:diffusionDrivenKeGamma} shows the enhanced effective diffusion coefficient as a function of $\gamma$ with  $\mathrm{Pe}_{p}=1$. As we expected, the enhanced effective diffusion coefficient is zero when $\gamma=0$, and converges to one as $\gamma$ increases to infinity.  This analysis shows that the dispersion of the stratified scalar can at most be doubled in the presence of diffusion-driven flow. In contrast, the effective diffusion coefficient of the passive scalar could be significantly enhanced by the diffusion-driven flow when the passive scalar diffusivity $\kappa_{p}$ is much smaller than the stratified scalar diffusivity $\kappa_{s}$, namely, $\mathrm{Pe}_{p}\gg \mathrm{Pe}_{s}$.

Second, we consider the dependence of effective diffusion coefficient on two different P\'eclet numbers.   The shear flow enhanced effective diffusion coefficient of a passive scalar is proportional to the square of the P\'eclet number $\mathrm{Pe}_{p}$, which has been demonstrated by many methods such as homogenization theory \cite{wu2014approach,camassa2010exact}, Aris moment approach \cite{aris1956dispersion,vedel2014time,vedel2012transient,ding2021enhanced}. All formulae of the effective diffusion coefficient \eqref{eq:Time depedent effective diffusivity diffusion driven flow parallel-plate channel}, \eqref{eq:Steady part Time depedent effective diffusivity diffusion driven flow parallel-plate channel} and \eqref{eq:effective diffusivity diffusion driven flow parallel-plate channel} are consistent with this conclusion. 
In contrast, the effective diffusion has a much more complicated dependence upon the stratified scalar's Peclet number, $\mathrm{Pe}_{s}$, as is clear from the formula given in \eqref{eq:effective diffusivity diffusion driven flow parallel-plate channel}.

\begin{figure}
  \centering
    \subfigure[$\gamma_{1}=1$]{
  \includegraphics[width=0.46\linewidth]{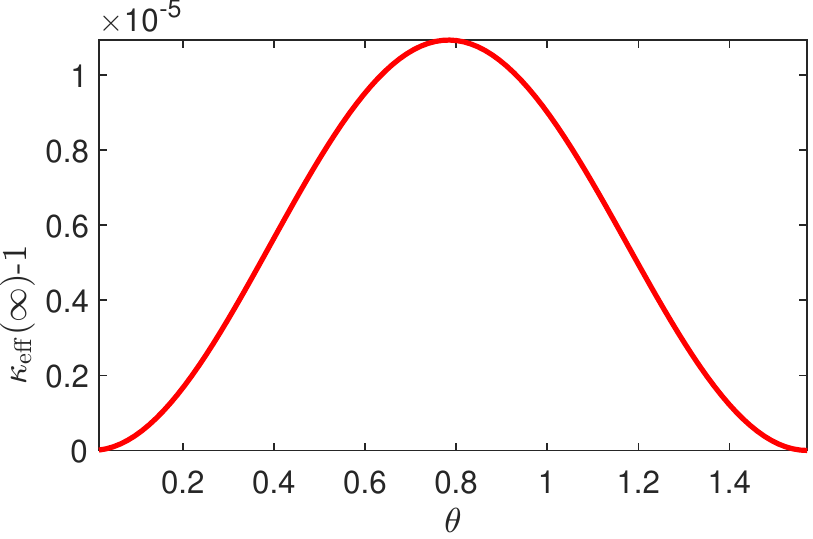}
}
    \subfigure[$\gamma_{1}=10$]{
  \includegraphics[width=0.46\linewidth]{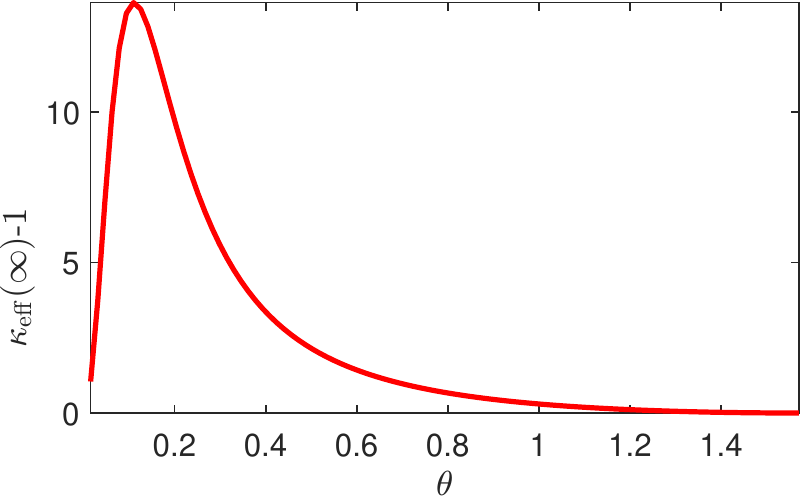}
}
  \hfill
  \caption[Enhanced effective diffusivity against the inclination angle ]
  {$\kappa_{\mathrm{eff}}-1$ against the inclination angle $\theta$. The parameters are $\mathrm{Pe}_{p}=\mathrm{Pe}_{s}=1$. In panel (a), $\gamma_{1}=1$, in panel (b) $\gamma_{1}=10$. }
  \label{fig:diffusionDrivenKeTheta}
\end{figure}

\begin{figure}
  \centering
    \includegraphics[width=0.46\linewidth]{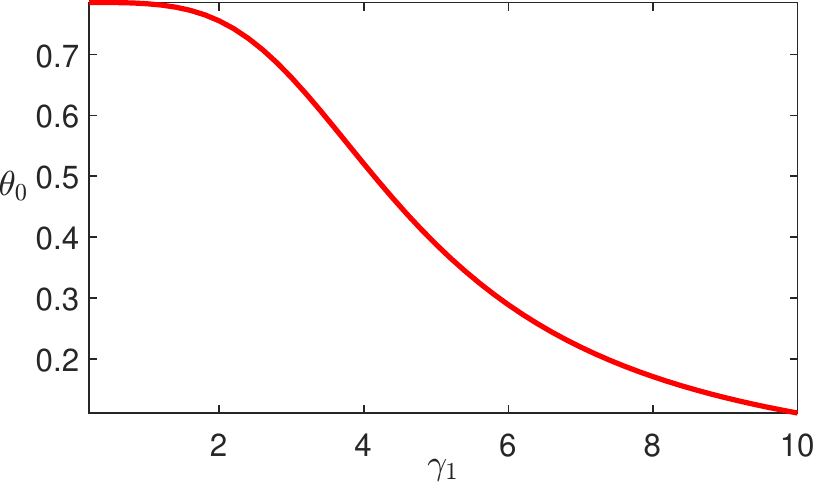}
  \hfill
  \caption[The optimal inclination angle for inducing the maximum effective diffusivity ]
  {The optimal inclination angle $\theta_{0}$ for inducing the maximum effective diffusion coefficient as a function of the parameter $\gamma_{1}=\frac{ 1}{\sqrt{2}}\left( \frac{\mathrm{Re} \mathrm{Pe}_{s}\Gamma_0 }{\mathrm{Fr}^2} \right)^{\frac{1}{4}}$. The parameters are $\mathrm{Pe}_{p}=\mathrm{Pe}_{s}=1$. }
  \label{fig:diffusionDrivenkeMaxTheta}
\end{figure}

Third, we study the dependence of the effective diffusion coefficient on the inclination angle. For fixed P\'eclet numbers and $\gamma_{1}=\frac{ 1}{\sqrt{2}}\left( \frac{\mathrm{Re} \mathrm{Pe}_{s}\Gamma_0 }{\mathrm{Fr}^2} \right)^{\frac{1}{4}}$, we have
\begin{equation}
\begin{aligned}
& \kappa_{\mathrm{eff}} (\infty) = 1+\frac{\mathrm{Pe}_{p}^{2}}{\mathrm{Pe}_{s}^{2}} \left(\frac{\gamma _1^8 \theta ^2}{22680}+\mathcal{O}\left(\theta ^{5/2}\right) \right),\quad \theta\rightarrow 0,\\
& \kappa_{\mathrm{eff}} (\infty) = 1+\frac{\mathrm{Pe}_{p}^{2}\left(\theta -\frac{\pi }{2}\right)^2}{2 \gamma _1 \mathrm{Pe}_{s}^{2}} \left( \frac{5}{2} \sin \left(2 \gamma _1\right)+5 \cos \left(\gamma _1\right) \sinh \left(\gamma _1\right)-5 \cosh \left(\gamma _1\right) \left(\sin \left(\gamma _1\right)+\sinh \left(\gamma _1\right)\right)  \right.\\
& \left. 6 \gamma _1 \sin \left(\gamma _1\right) \sinh \left(\gamma _1\right)+\gamma _1 +\left(\cosh \left(2 \gamma _1\right)-\cos \left(2 \gamma _1\right)\right) \right)+\mathcal{O}\left( \theta- \frac{\pi}{2} \right)^{3} ,\quad \theta \rightarrow \frac{\pi}{2}.
\end{aligned}
\end{equation}

\begin{figure}
  \centering
    \includegraphics[width=0.46\linewidth]{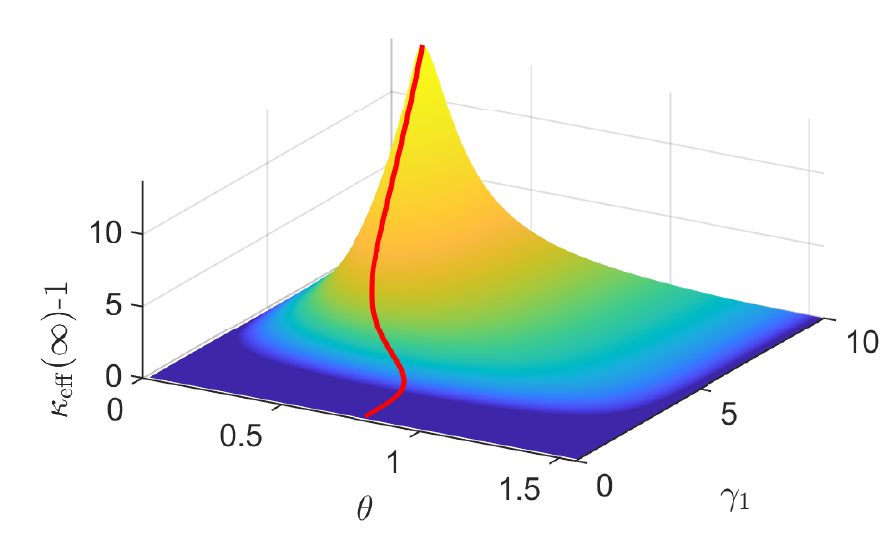}
  \hfill
  \caption[The enhanced effective diffusivity as a function of the inclination angle and the parameter $\gamma_{1}$]
  {The enhanced effective diffusion coefficient as a function of the inclination angle and the parameter $\gamma_{1}$. The red solid line indicates the optimal inclination angle $\theta_{0}$ for inducing the maximum effective diffusion coefficient when the parameter $\gamma_{1}=\frac{ 1}{\sqrt{2}}\left( \frac{\mathrm{Re} \mathrm{Pe}_{s}\Gamma_0 }{\mathrm{Fr}^2} \right)^{\frac{1}{4}}$ is given. The parameters are $\mathrm{Pe}_{p}=\mathrm{Pe}_{s}=1$. }
  \label{fig:diffusionDrivenkeThetaGamma}
\end{figure}

Figure \ref{fig:diffusionDrivenKeTheta} plots the enhanced effective diffusion coefficient as a function of the inclination angle $\theta$.
% Since $\cot ^{2} (\theta)$ and $\sin^{2} (\theta)$ are symmetric with respect to $\theta= \frac{\pi}{4}$, the curve of enhanced effective diffusion coefficient is also symmetric with respect to $\theta=\frac{\pi}{4}$.
The enhanced effective diffusion coefficient vanishes at $\theta=0$ and $\frac{\pi}{2}$, which is consistent with the asymptotic expansions \eqref{eq:effective diffusion coefficient gamma limit}. The shape of this curve depends on the value of $\gamma_1$. It is symmetric when $\gamma_{1}$ is small, and skewed when $\gamma_{1}$ is large.  Numerical calculation shows that the enhanced effective diffusion coefficient reaches the maximum value $\kappa_{\mathrm{eff}}\approx 0.0000109356$ at $\theta\approx 0.783409<\frac{\pi}{4}$ when $\gamma_{1}=1$, and reaches the maximum value $\kappa_{\mathrm{eff}}\approx 13.6319$ at $\theta\approx 0.110802 $ when $\gamma_{1}=10$ . Figure \ref{fig:diffusionDrivenkeMaxTheta} shows the optimal inclination angle $\theta_{0}$ for inducing the maximum effective diffusion coefficient as a function of the parameter $\gamma_{1}$. For small $\gamma_{1}$, the optimal inclination angle is around $\theta= \frac{\pi}{4}\approx 0.785398$ which can be seen from equation \eqref{eq:effective diffusion coefficient gamma limit}. As $\gamma_{1}$ increases, the optimal inclination angle decreases. The dependence of the enhanced effective diffusion coefficient on the inclination angle and the parameter $\gamma_{1}$ is summarized in figure \ref{fig:diffusionDrivenkeThetaGamma}.

\section{Conclusion and discussion}
Here we have explored the diffusion-driven flow in  the tilted parallel-plate channel domain with a linear density stratification along with the effective mixing of a diffusing passive scalar advected by this flow. Exact expressions for the flow were derived using an  eigenfunction expansion, where it was established that the unsteady flow converges to the steady solution monotonically or oscillatory depending on the relation between the Schmidt number and the non-dimensionalized diffusivity. We demonstrated that when $\kappa_{2}= \mathrm{Sc}$, we have the most observable oscillations in the flow evolution.  We then calculated the exact scalar distribution variance evolution and effective diffusion coefficient for the passive scalar. The formula demonstrates that the diffusion-driven flow could significantly enhance the effective diffusion coefficient of the scalar, especially when the molecular diffusivity of the passive scalar is much smaller than the stratified scalar diffusivity. This enhancement could have potential applications in geophysics and in microfluidics.  We discovered a nonlinear dependence between the enhanced effective diffusion coefficient and the P\'eclet number of the stratified scalar, which is different from the typical quadratic scaling relation for the passive scalar in a shear flow. 

Future work includes several directions. First, the steady diffusion-driven flow has been studied in many different boundary geometries \cite{grayer2020dynamics,page2011steady,page2011combined,french2017diffusion}. We are interested in investigating the time-dependent solution in those domains. Second, the current analysis assumes a linear stratification to simplify the calculation. In future work, we are interested in analyzing the flow and scalar evolution using full numerical simulations to further explore the validity of the Boussinesq approximation. Third, the diffusion-driven flow might exist in the presence of other external force fields as long as the direction of the external force field is not parallel to the impermeable boundary. One possible external force field is the electric field, and therefore we expect the diffusion-driven flow could be observed in some electrohydrodynamic problems.

\section{Acknowledgements}
We acknowledge funding received from the following National Science Foundation Grant Nos.:DMS-1910824; and Office of Naval Research  Grant No: ONR N00014-18-1-2490.

\appendix

\section{Reduction of the effective diffusion coefficient}

We present a simple shear flow that explicitly demonstrates a case in which the dynamic effective coefficient can be less than one on transient timescales.
When $v (y,t)=\sqrt{2}\cos \pi y \left( e^{-2t}-e^{-t} \right)$, the solution of equation \eqref{eq:Aris moment 1 diffusion-driven flow} is 
\begin{equation}
\begin{aligned}
&T_{1}=\frac{e^{-\pi ^2 t} \left(e^{\left(\pi ^2-2\right) t} \left(\left(\pi ^2-2\right) \left(-e^t\right)+\pi ^2-1\right)-1\right)}{2-3 \pi ^2+\pi ^4}, \\
\end{aligned}
\end{equation}

The effective diffusion coefficient is given by
\begin{equation}
\begin{aligned}
\kappa_{\mathrm{eff}} (t) =1- \mathrm{Pe}_{p}^{2}\frac{e^{-\left(2+\pi ^2\right) t} \left(e^t-1\right) \left(e^{\left(\pi ^2-2\right) t} \left(\left(\pi ^2-2\right) e^t+\pi ^2-1\right)-2 \pi ^2+3\right)}{2-3 \pi ^2+\pi ^4}.
\end{aligned}
\end{equation}
When $\mathrm{Pe}_{p}=1$,   $\kappa_{\mathrm{eff}} (1)\approx 0.986359<1$, namely, the longitudinal dispersion is temporally reduced by this time-dependent shear flow.

\bibliographystyle{elsarticle-harv}

\end{document}